\documentclass{aa}
\def\sqr#1#2{{\vcenter{\vbox{\hrule height.#2pt
    \hbox{\vrule width.#2pt height#1pt \kern#1pt \vrule width.#2pt}
    \hrule height.#2pt}}}}

\def\bull{\vrule height .9ex width .8ex depth -.1ex}
\def\simlt{\lower.5ex\hbox{$\; \buildrel < \over \sim \;$}}
\def\simgt{\lower.5ex\hbox{$\; \buildrel > \over \sim \;$}}
\def\lesssim{\lower.5ex\hbox{$\; \buildrel < \over \sim \;$}}
\def\msun{$\rm M_\odot$}
\def\teff{$T_{\rm eff}$}
\def\tkh{$\rm t_{\rm KH}$}

\usepackage{graphicx}
\begin{document}
   \title{Non-gray rotating stellar models and the evolutionary history of the 
Orion Nebular Cluster}

%   \subtitle{}

   \author{N.R. Landin\inst{1,2}, P. Ventura\inst{2}, F. D'Antona\inst{2}, 
      L.T.S. Mendes\inst{1,3} \and L.P.R. Vaz\inst{1} 
          }

   \offprints{N.R.Landin}

   \institute{Depto.\ de F\'{\i}sica,
              Universidade Federal de Minas Gerais, C.P.702, 30161-970 --
              Belo Horizonte, MG, Brazil; \\
              \email{nlandin@fisica.ufmg.br, lpv@fisica.ufmg.br}
              \and
              Osservatorio Astronomico di Roma
              Via Frascati 33 00040 MontePorzio Catone -- Italy\\
              \email{ventura@mporzio.astro.it, dantona@mporzio.astro.it}
              \and
              Depto.\ de Engenharia Eletr\^onica,
              Universidade Federal de Minas Gerais, C.P.702, 30161-970 --
              Belo Horizonte, MG, Brazil; \\
              \email{luizt@cpdee.ufmg.br}  
             }

   \date{Received \dots; accepted \dots}

% \abstract{}{}{}{}{} 
% 5 {} token are mandatory
 
\abstract
{Rotational evolution in the pre-main sequence is described with 
new sets of pre-MS evolutionary tracks including rotation, non-gray boundary 
conditions (BCs) and either low (LCE) or high convection 
efficiency (HCE). }
{Using observational data and our theoretical predictions, we aim at
constraining
(1) the differences obtained for the rotational evolution of stars within
the ONC by means of these different sets of new models; 
(2) the initial angular momentum of low mass stars,
by means of their templates in the ONC.
}
{We discuss the reliability of current stellar models for the pre-MS. 
While the 2D radiation hydrodynamic simulations predict HCE in pre-MS,
semi-empirical calibrations either seem to require that convection is less efficient 
in pre-MS
than in the following MS phase (lithium depletion) or are still contradictory 
(binary masses).
We derive stellar masses and ages for the ONC by using both LCE and HCE.
}
{The resulting mass distribution for the bulk of the ONC population is in the 
range 0.2$-$0.4{\msun} for our new non-gray models and,  as in previous analyse,
in the range 0.1$-$0.3{\msun} for models having gray BCs.
In agreement 
with Herbst et al.\ (2002) we find that a large percentage ($\sim$70\%) of 
low-mass stars (M\simlt0.5{\msun} for LCE; M\simlt0.35{\msun} for HCE) in the
ONC appears to be fast rotators (P$<$4days). Three possibilities are open: 1) 
$\sim$70\% of the ONC low mass stars lose their disk at 
early evolutionary phases; 2)their ``locking period'' is shorter; 
3) the period evolution is linked to a different morphology of the 
magnetic fields of the two groups of stars. We also estimate the range 
of initial angular momentum consistent with the observed periods.  
}
{The comparisons made indicate that a 
second parameter is needed to describe convection in the pre-MS, possibly 
related to the structural effect of a dynamo magnetic field.}
%{Place here some concluding remarks?\dots}


\authorrunning {Landin et al.}
%\titlerunning {Non-gray rotating stellar models and the evolutionary history of the ONC}
\titlerunning {Non-gray rotating stellar models and the evolution in the ONC}

\maketitle

%________________________________________________________________

\section{Introduction}

We compute new sets of pre-MS models with the ATON code 
for stellar evolution, the version presented in Mendes et al. (\cite{mendes}), which
includes stellar rotation according to the description by Endal \& Sofia 
(\cite{endal}), updated for the present work to employ non-gray 
boundary conditions by Allard \& Hauschildt (\cite{allard1}) and  Allard 
et al.\ (\cite{allard00}). 

%{\bf Our aim is to understand which are the appropriate physical constraints to be 
%used for a general description of the evolution of stellar angular momentum, 
%in particular the choice of the initial value of the stellar 
%angular momentum and its variation with time. We also aim to understand how the 
%convection efficiency behaves during the 
%pre--MS phase.}
Our main goal is to improve our understanding of the appropriate
physical constraints to be used for a general description of the
evolution of stellar structure and its angular momentum with time. We
are particularly interested on the choice of the stellar initial angular
momentum and its variation with time, and, also, on the importance of
convection efficiency during the pre-main sequence.  
To do this, we check our choices with respect to sets of relevant 
observations.

Since the pioneering work by Henyey et al. (\cite{henyey}) and Hayashi 
(\cite{hayashi}) it is commonly accepted that pre-MS stars derive their 
luminosity by gravitational contraction, with the only exception of the short 
deuterium burning phase. Derivation of masses of young stellar associations has 
then generally been made by standard hydrostatic stellar models including 
deuterium burning, an approach that we will use in this paper. This 
procedure rests on the assumption that neither the residual accretion after the 
protostellar phase nor the uncertainty in the zero point of ages affect 
the results in a strong way.

Nevertheless, the theoretical description of moderately low and low mass 
objects is affected by the first principle uncertainties in the description of 
some physical inputs, in particular opacities, convection, equation of state 
(D'Antona \cite{franca5}) and treatment of boundary conditions (BCs, Chabrier 
\& Baraffe \cite{chabrier}). Since the low mass stars in the pre-MS are fully 
convective and over-adiabatic, any change in the convective model substantially 
alters the location of the track in the theoretical HR plane. The use of a less 
efficient treatment of convection leads to larger temperature gradients, so 
that, for a given luminosity, the structure readjusts on a more expanded 
configuration, with a consequent shift of the track to lower effective 
temperatures ($T_{\rm eff}$) (e.g. D'Antona \& Mazzitelli \cite{franca1}, 
\cite{franca3}; D'Antona \& Montalb\'an \cite{franca2}). The path followed by 
the theoretical pre-MS tracks on the HR diagram is also dependent on the 
boundary conditions used to fit the numerical integration of the structural 
equations of the interior with the atmosphere. The use of a non-gray 
atmospheric treatment shifts the tracks to cooler $T_{\rm eff}$s within an 
extended interval of masses and ages (Montalb\'an et al.\ \cite{josefina}). The 
effect due to the non-grayness of the atmosphere is, in many cases, overwhelmed 
by the uncertainties related to the treatment of convection, that has a 
similar, but even stronger effect on the tracks, with only a few exceptions 
that are relevant for this work, as will be discussed in \S
\ref{ngconv}. 

Young stellar clusters provide a unique opportunity to test stellar pre%
-MS models. Many studies in the past have been focused on the Orion Nebula 
Cluster (ONC), because it contains thousands of pre-MS objects. Hillenbrand 
(\cite{hillen97}) measured the V and I magnitudes (and colors) of $\sim$900 
stars and located them in the theoretical HR diagram by using bolometric 
corrections and taking into account the interstellar extinction for the 
determination of colors and bolometric magnitudes. In spite of the non-%
negligible uncertainties weighting on the derivation of these stellar 
parameters (Hillenbrand et al. \cite{hillen97}), such a work has been widely used, in 
connection with theoretical pre-MS tracks, to infer important information 
concerning the cluster itself. The mass and age distributions and the slope of 
the mass function can all be estimated by inferring the appropriate mass and 
age for each observed star. It is clear that the results obtained with this 
approach will depend to a certain extent on the set of tracks used to perform 
the analysis, and on the physical inputs adopted to calculate the evolution. 
This is confirmed by the fact that, specifically for the ONC, studies that used 
different sets of evolutionary tracks reached significantly different 
conclusions, particularly with respect to the mass and age distribution, and, 
more important, to the age spread and, consequently, to the evolutionary 
history of the star formation process within the cluster (Palla \& Stahler 
\cite{palla3}).

In the past few years, new detailed observational studies of the ONC have been 
undertaken, focused on the rotational properties of the stars. Stassun et al.\ 
(\cite{stassun}) and Herbst et al.\ (\cite{herbst02}) measured the rotational 
periods of $\sim$400 stars belonging to the ONC. All of them are in the 
Hillenbrand (\cite{hillen97}) sample, so they can be located on the HR diagram. 
More recently, Stassun et al.\ (\cite{stassun04}) and Flaccomio et al.\ 
(\cite{flacco03a}, \cite{flacco03b}) reanalyzed all the archival {\it Chandra} 
observations of the ONC studying in great detail the X-ray properties of the 
observed objects, in an attempt to elucidate the origin of X-ray emission in 
pre-MS stars. All this available information renders the ONC an excellent 
laboratory to test stellar evolution theories of the pre-MS phase.
 
None of the previous analyses of the ONC rotational database has been done  
using non-gray models, and the effect of using different convection efficiencies
has not been extensively tested. We therefore decided to use this database to 
test and calibrate our new sets of non-gray tracks for rotating stellar models, 
with masses in the range 0.085$\leq$$M/M_{\odot}$$\leq$3.8. As a byproduct, we 
will have insight on how much the rotational properties of this population depend
on the choice of the evolutionary tracks. At present, there 
is no definitive observational constraint that can be used to choose among the 
tracks obtained by using different convection inputs. In this work we derive 
masses and ages with different sets, in order to appreciate how much the 
results we are presently interested in, namely those on the angular momentum 
evolution, depend on the choice. In particular, we will test the role of
(a) boundary conditions (gray and non-gray),
(b) different convection efficiencies in the framework of the MLT, and
(c) rotating and non-rotating models.

As indicated by the non-negligible differences among the set of tracks adopted, we 
show that the detailed period distribution as a function of mass and age is 
dependent on the physical inputs. 
However, the qualitative information on the 
rotational distribution of stars of different mass in the ONC remains similar, 
and we confirm that the distribution of periods is bimodal only for masses 
larger than a ``transition" mass depending on the convection model.

In \S \ref{sect2} we summarize the uncertainties in the theoretical models, and also try 
to clarify some issues left ambiguous in the recent literature on 
pre-MS evolution. In \S \ref{models} we provide a brief description of the latest 
rotational, non-gray version of the ATON stellar evolution code, and describe 
the choice of the main physical and chemical inputs used for the present work
in \S \ref{input}. 
The role played by the convective modeling on the location 
of pre-MS tracks on the HR diagram, the role played by non-grayness, and the 
lithium pre-MS depletion are discussed in \S \ref{sect5}. In \S \ref{sect6} we present
the photometric and rotational data related to the ONC stellar population,
and the derivation of masses and ages from our new sets of tracks.
The results on the rotation period distribution and its possible interpretation(s) 
are discussed in \S \ref{poponc}. Conclusions are given in \S \ref{sect8}.

%______________________________________________
\section{Uncertainties in the standard hydrostatic stellar models}
\label{sect2}

We briefly rewiew the reliability of the procedure of deriving ages and 
masses of young stellar population based on hydrostatic evolutionary tracks. We 
first discuss two preliminary important physical points: (1) Is it appropriate to
neglect residual accretion from the disk? (2) As the models disregard the
hydrodynamic formation process, how do we take into account the zero point of ages?
We summarize the well known uncertainties in the location
of evolutionary tracks in the HR diagram and show why current data 
do not allow us to choose among models.

\subsection{The role of residual accretion}
The pre-MS phase begins after the main accretion phase, during which the 
protostellar core is formed and the star is embedded into the dust of the 
forming cloud. When objects of low mass become luminous in the visible or near 
infrared bands, the main accretion phase can be considered finished, although 
the accretion disk is still present in many cases. Accretion then will not 
change the final stellar mass much, while the presence of the accretion disk 
may be very important in determining the rotational evolution of the star.
In the pre-MS, accretion still may occur 
at rates in the range $\sim$3$\times 10^{-9} - 4$$\times$$10^{-7}$\msun/yr 
(Basri \& Bertout, \cite{basri89}) for classical T~Tauri (TT) star, 
and mostly below $\sim$$10^{-10}$\msun/yr for very low mass stars 
below $\sim$0.2{\msun} (Mohanty et al. 
\cite{mohanty05}). Walter (\cite{walter}) and Strom et al.\ (\cite{strom1989}) 
noticed the coexistence of classical and weak TT
stars in the same region of the HR diagram, corresponding to an age range 
(derived from standard hydrostatic models without accretion) from 
$\sim$3$\times 10^5$yr to $\sim$3$\times 10^6$yr, concluding that the disk 
evolution is decoupled from the evolution of the central star, and that the 
disk may disappear either early or at late stages. This will affect the 
stellar rotation in different ways, as shown by subsequent studies (e.g. 
Bouvier et al. \cite{bouvier1993}), interpreted in terms of magnetic disk 
locking (Bouvier et al. \cite{bouvier1997}).

Can the residual mass accretion alter the stellar evolution? Deuterium 
burning from the accreted matter provides a luminosity
$L$$\sim$$15L_\odot$$\times$$(\dot{M}/(10^{- 5}$\msun/yr))
(Stahler \cite{stahler1988}), which could 
influence the evolution only for very high rates of mass accretion, or at very 
low luminosity. Accretion itself may alter the evolution if the accretion 
timescale $M/\dot{M}_{\rm crit}$\ is of the order of the thermal (Kelvin-Helmholtz) 
timescale \tkh. At high luminosity, where \tkh$\sim$$10^5$yr for a typical 
TT, this constraint is not respected only for very high accretion rates. At 
about the solar luminosity \tkh$\sim$2$\times$$10^6$yr, for which 
$\dot{M}_{\rm crit}$$\sim$5$\times$$10^{-7}$\msun/yr, which is at the upper boundary
of the values observed in TT. In this case, or in similar cases, the 
evolution can be altered, but this $\dot{M}$ cannot be sustained 
for a very long time, otherwise we would not find low masses 
surviving. Thus we conclude that evolution in which accretion plays a dominant 
role, such as described e.g. by Hartmann et al. (\cite{hartmann1997}) and Tout 
et al. (\cite{tout1999}) may be relevant only for a short lifetime, in a small 
fraction of the pre-MS stars, and that the bulk properties of young stellar 
associations can be derived by adopting traditional pre-MS hydrostatic models.

\subsection{The zero point of stellar ages}
One other item that must be taken into account is the definition of the zero 
point of ages, which, for the hydrostatic pre-MS evolution, is connected to 
the location in luminosity of the ``starting model", i.e. the internal 
thermodynamic conditions within the star when the mass accretion process ends. 
The uncertainty in the starting models is reflected in the age-luminosity 
relationship. A first uncertainty is due to the fact that we do not know 
whether the deuterium burning phase occurs during the visible pre-MS or during 
the main accretion phase: this depends on the average protostellar accretion 
rate (see D'Antona \& Mazzitelli \cite{franca3} for a discussion). For the 
stars that do 
not completely burn deuterium in the protostellar phase (M$\lesssim 0.5$\msun), 
the typical uncertainty in the lifetime, however, must be of the order of \tkh\ 
at the end of the main accretion phase (e.g. Tout et al. \cite{tout1999}). 
In the model 
by Palla \& Stahler (\cite{palla1}, \cite{palla2}), in which the stellar core 
of low mass stars emerges at visible wavelengths close to the deuterium burning 
region in the HR diagram (birthline) due to the ``thermostat" action of D-%
burning, the typical \tkh\ is the thermal timescale at the D-burning 
luminosity and radius, $\sim$$10^5$yr for masses of a few tenths of \msun, and up 
to $\sim$$10^6$yr for 0.1\msun.  For the lowest masses, however, the 
``birthline" concept may not be valid, as the main accretion phase may end 
before deuterium is ignited, and the stars probably start their hydrostatic 
evolution at higher luminosities, where the thermal timescale, and thus the 
uncertainty in the zero point of ages, would be smaller.
\par 
On the other hand, we must be careful not to confuse the ``numerical" 
uncertainties with the ``physical" uncertainties. Hydrostatic contraction will 
let the star ignite D-burning as soon as the central physical conditions allow 
it, and it seems very unlikely that we can take as a zero point of the pre-MS a 
model with central temperatures in the middle of the deuterium burning phase, 
disregarding the action of the thermostat\footnote{Some models of the widely 
adopted grids by Baraffe et al. (\cite{baraffe}) begin their evolution at a luminosity 
in the middle of D-burning, which lasts for $\sim$2$\times$$10^6$yr at 
0.1\msun. This is due to the fact that the lowest available gravity in the 
NextGen (Allard \& Hauschildt \cite{allard1}, AH) atmospheric grid, which these
models employ, was $\log g=$3.5. The more 
recent models by Baraffe et al. (\cite{baraffe2002}) start at lower gravity, 
and reach the D-burning luminosity by contracting in thermal equilibrium. Of 
course, in these two sets of models the temporal 
evolution of luminosity is very different for the first $\sim$2$\times$$10^6$yr, but
this does not automatically mean that this figure is the physical uncertainty in the 
ages, as the ``numerical" choice of starting from a model with central 
temperature in the middle of the D-burning stage
does not necessarily correspond to a physically plausible behavior 
(Montalb\'an \& D'Antona \cite{md2006}).}.

\subsection{The role of non-gray atmospheres}
In the gray atmosphere approximation the increase of the temperature T from the
surface moving inwards is described via a relationship between T and
the optical depth ($\tau$), and the pressure is calculated by integrating the
hydrostatic equilibrium equation. The match between the interior
and the external layers is made at $\tau$=2/3.  
Any link between pressure and temperature in the atmosphere, fixed by 
the either a convective or radiative gradient, is ignored. In the
non-gray treatment a self-consistent integration is performed down to
an optical depth at which the diffusion approximation is valid 
(Morel et al.\ \cite{morel}), and including the treatment of atmospheric 
convection, which cannot be neglected at low $T_{\rm eff}$s. The use of
frequency-dependent opacities may also modify the onset of convection
within the atmosphere, and both the $T_{\rm eff}$ and the colors of the tracks
can be strongly affected.
The necessity of adopting outer boundary conditions based on realistic non-gray 
atmosphere models for the pre-MS and low mass MS
was pointed out by Chabrier \& Baraffe (\cite{chabrier}; and references 
therein), who have shown that the use of radiative T($\tau$) relations or gray 
atmosphere models is invalid when molecules form near the photosphere, at 
\teff\ below 4000K. Outer boundary conditions based on the gray assumptions 
yield hotter models for a given mass. According to Baraffe et al. 
(\cite{baraffe}) the 
use of an inappropriate outer boundary condition, such as the Eddington 
approximation, yields an overestimation of \teff\ for a given mass 
up to 300 K. 

\subsection{The role of convection} \label{convrole}
There is broad consensus in the literature that the treatment of 
superadiabatic convection in the pre-MS affects the tracks location to
a great extent.
In a series of works focusing on understanding the role played by different
physical inputs on the pre-MS evolution, D'Antona \& Mazzitelli 
(\cite{franca3}), D'Antona \& Montalb\'an (\cite{franca2}), and Montalb\'an et 
al.\ (\cite{josefina}) outlined the major impact of convection modeling on the 
location of the stellar tracks in the HR diagram. Convection was found to be by 
far the most relevant ingredient influencing the determination of the mass and age 
of observed stars.

When modeling convection, it is essential to specify the mixing scale 
$\Lambda$, i.e. the typical distance that convective eddies travel before 
dissolving and delivering their excess gravo-thermal heat to the environment. 
The role played by $\Lambda$ is relevant for the determination of the 
temperature gradient, as the conservation of flux implies that a larger 
$\Lambda$ must be compensated by a lower degree of overadiabaticity. The 
convective flux behaves as $F_C$$\sim$$\Lambda^2$ in zones where most of the 
energy is carried by convection. Conversely, within low efficiency convective 
regions, $F_C$$\sim$$\Lambda^8$. The impact of the choice of $\Lambda$ is 
therefore more evident where convection is not efficient. The pre-MS tracks 
result to be particularly sensitive to convective modeling, because the 
surface convection extends to most of the (if not the whole) star, and the low 
densities involved (particularly in the early phases of gravitational 
contraction) make the convective process highly inefficient.

Presently, the main ways of computing convection in stellar envelopes,
for wide grids of stellar models, are:
\begin{enumerate}
  \item{The traditional mixing length theory --- MLT (B\"ohm-Vitense \cite{bohm}, and subsequent 
variations of this same model) --- assumes that
both the dimension of the convective eddies and the mixing length are
proportional to the local value of the pressure scale height, i.e. 
$\Lambda=\alpha H_p$, where $\alpha$ is a free parameter that is usually 
calibrated in order to reproduce the solar radius.}
  \item{In the Full Spectrum of Turbulence model --- FST (Canuto et al.\ \cite{canuto1}) --- 
the whole spectrum of eddies' dimensions is considered, and the mixing length is taken as 
the distance of the nearest convective border.}
  \item MLT, in which the $\alpha$\ value for each gravity and \teff\
  is calibrated upon 2D or 3D hydrodynamical simulations.
\end{enumerate}
Ludwig et al. (\cite{ludwig}), using their 2D radiation hydrodynamic models,
have provided a calibration of the parameter 
$\alpha$ in a wide region of \teff's and gravities\footnote{$\alpha$ is mapped in the domain 
$T_{\rm eff} = 4300 - 7100$~K, $\log g = 2.54 - 4.74$} to be used in the computation of 
{\em gray} stellar models. These 2D models indicate that convection in the
pre-MS is on average more `efficient' than in the MS, corresponding to a larger $\alpha$.
The idea of calibrating the average $\alpha$\ using numerical simulations has been extended 
by now to  a  few 3D computations:
Ludwig et al.\ (AHS, \cite{ludwig2002}), for an M dwarf at $T_{\rm eff} = 2800$~K and $\log g =5$,
find  $\alpha \simeq$2.1; and Trampedach et al.~(\cite{tramp}),
for the range of main sequence gravities  and  $\log T_{\rm eff}$=3.68--3.83, find  
$\alpha \simeq$1.6--1.8 in the whole range.
Asplund et al.~(\cite{asplund00}) have compared 2D and 3D atmosphere models for Sun,
and found that  the 2D solar model  has marginally larger gradients than the 3D one. 
Although an extrapolation to regions not explicitly computed is not allowed, 
also these few 3D models indicate efficient convection in the overadiabatic 
%envelope. Montalb\'an \& D'Antona (\cite{dm2005}) and Montalb\'an \& D'Antona 
%(in preparation) have computed gray models by using this calibration: in fact, 
envelope. Montalb\'an \& D'Antona (\cite{md2006})  
have computed gray models by using this calibration:  
the tracks they obtain are very similar to the FST tracks\footnote{
The FST convection model corresponds to {\em very efficient convection}, as 
shown by the quasi-coincidence of the resulting tracks with the tracks 
employing the MLT $\alpha$ calibrated on the 2D hydrodynamic models. 
We find confusing and misleading the statement by Baraffe et 
al. (\cite{baraffe2002}), who point out that the FST model is very {\em 
inefficient} in the upper solar layers, where it provides results that are not 
consistent with the hydrodynamic simulations of convection for the solar model 
(nevertheless, remember that the FST model provides a better fit than MLT for 
the spectrum of solar oscillations, see e.g. Canuto \& Christensen-Dalsgaard, 
\cite{cd1998}). What matters in the description of the pre-MS is the average efficiency 
of convection in the {\em whole} superadiabatic envelope, and this is very 
large for the FST, roughly corresponding to an $\alpha$\ value in the MLT 
description somewhat larger than 2.}. Unfortunately, a very efficient 
convection in pre-MS is {\em not} consistent with the lithium 
depletion patterns of young open clusters 
(D'Antona \& Montalb\'an \cite{franca2}). 
Any attempt to calibrate convection efficiency in pre--MS
by means of comparisons between binary masses dynamically determined and those
assigned from different sets of evolutionary tracks (HCE and LCE), seems to 
be ambiguous (Landin et al.\ \cite{landin06}).

\subsection{The role of convection coupled with non-gray atmospheres}
We note a point that is generally overlooked, but that is crucial in
order to understand the relevant parameters in track building: 
Montalb\'an et al.\ (\cite{josefina}) have shown that the problem of convection 
is intertwined in a subtle way with the problem of non-gray boundary 
conditions. For example, in the MLT framework, convection in the pre-MS should be 
described not only by the ratio $\alpha=l/H_p$\ in the interior of the star 
($\alpha_{\rm in}$), below the non-gray atmosphere, but also by the value that 
this parameter has in the atmosphere itself ($\alpha_{\rm atm}$), and by the 
matching point between atmosphere and interior ($\tau_{\rm ph}$).
In particular, the two widely used sets of model atmosphere by Baraffe et al. 
(\cite{baraffe}) are referred to as the set built with $\alpha$=1.9 and the set 
having $\alpha$=1.0. Thus they are supposed to provide a clue to how the tracks 
vary by changing from a moderately high convective efficiency (HCE) represented 
by the set $\alpha$=1.9, to a low convection efficiency (LCE) represented by 
the set $\alpha$=1. However, in Baraffe et al. (\cite{baraffe}),
the parameter $\alpha$ refers only 
to the value of $\alpha_{\rm in}$, and is misleading for two reasons: (1) the set 
$\alpha_{\rm in}$=1.9 stops at masses M$\ge$0.6\msun, and only the set 
$\alpha_{\rm in}$=1.0 is available for masses M$<$0.6\msun; (2) for both sets, the
atmospheric model grid adopted is the same, and computed with 
$\alpha_{\rm atm}=1.0$. Montalb\'an et al.\ (\cite{josefina}) have shown that the 
fact that most of the superadiabatic part of the envelope is computed 
with a very inefficient convection ($\alpha_{\rm atm}$=1.0) shifts the \teff\ by
$\sim -150$K for the solar pre-MS. Thus one may be led to attribute the 
smaller \teff s of the Baraffe et al. (\cite{baraffe}) tracks to the use of the non-%
gray atmospheres, whereas they are due in part to the fact that these non-gray 
atmospheres are computed with LCE. 

%__________________________________________________________________

\section{Rotating stellar models} \label{models}

The pre-MS tracks were calculated by means of the ATON stellar evolution code. 
A full and detailed description of the numerical structure can be found in 
Mazzitelli (\cite{italo1}) and Mazzitelli et al.\ (\cite{italo2}). Here we 
briefly recall the main micro- and macro-physics input.

\subsection{Rotation}
As described in Mendes et al.\ (\cite{mendes}) rotation was implemented in the 
ATON code according to the approach followed by Endal \& Sofia (\cite{endal}), 
which uses the Kippenhahn \& Thomas (\cite{kippen70}) method improved with a 
potential function that includes a term related to the  distortion of the 
figure of the star. In this approach the spherical surfaces used in standard 
stellar models are replaced by the equipotential surfaces. 

The current version of the code allows one to choose among three rotational schemes:
\begin{itemize}
\item{Rigid body rotation throughout the whole star}
\item{Local conservation of angular momentum in the whole star}
\item{Local conservation of angular momentum in radiative regions
plus rigid body rotation in convective zones}
\end{itemize}
It is possible to consider only the hydrostatic effects of rotation and/or 
to include also the internal angular momentum redistribution and surface 
angular momentum loss.

\subsection{Atmospheric treatments}

In addition to the gray atmospheric boundary conditions in the version by 
Mendes et al.\ (\cite{mendes}), the ATON2.4 code can now adopt non-gray 
atmospheric integration. In the first case the internal structure is matched at 
$\tau$=2/3 with the values of P and T found via a Krishna-Swamy 
(\cite{krishna}) $T(\tau)$ relation. In the non-gray case we can choose the 
optical depth at which the matching is done: typical values are in the range 
1$\le$$\tau$$\le$100.
For the atmospheric structure we can choose among several sets of models:
(1) either the MLT or the FST models by Heiter et al.\ (\cite{heiter}); 
(2) the AH+AHS grid, based on the NextGen models by  Allard \& Hauschildt (\cite{allard1}), 
complemented by the low gravity models by Allard et al.\ (\cite{allard00}).
To follow the evolutions starting from early, low-gravity stages
we merged the NextGen tables (only available for $\log g$$\ge$3.5) 
with the more recent low-gravity models (2$\le$$\log g$$\le$3.5). Eventually, we
obtained rectangular tables in the range 2$\le$$\log g$$\le$6 and 
2000~K$\le$$T_{\rm eff}$$\le$6800~K.

\subsection{Input micro-physics}
The radiative opacities are taken from Iglesias \& Rogers (IR93, \cite{rogers1}),
extended by the Alexander \& Ferguson (AF94, \cite{alexander}) tables in 
the low-temperature regime. The OPAL equation of state (Rogers et al.\ 
\cite{rogers2}, R96) is used in the range 3.7$<$$\log T$$<$8.7, while in the 
low-T high density regime we use the Mihalas et al.\ (M88, \cite{mihalas}) 
EOS. The nuclear network includes 14 elements and 22 reactions; the relevant 
cross-sections are taken from Caughlan \& Fowler (\cite{caughlan}).

\subsection{Convection}
The borders of the instability regions are found via the Schwarzschild 
criterion. The convective fluxes can be computed either by the FST model 
(Canuto et al.\ \cite{canuto1}), or by the MLT (B\"ohm-Vitense \cite{bohm}); in 
the latter case the free parameter $\alpha = \Lambda /H_p$ determining the 
mixing length can be arbitrarily selected. 

\section{Input of present models} \label{input}

\begin{table}[b]
\caption{Main physical parameters of the present models.} \label{modpar}
\centering
\begin{tabular}{ll}
\hline \hline
Parameter                 & Input                         \\ \hline
Mass                      & 0.085--3.8 M$_{\odot}$        \\
Boundary Conditions       & AH+AHS                        \\
Matching point            & $\tau$=10                     \\
Convection model          & MLT                           \\
MLT parameter             & $\alpha$=1.0, 2.0 and 2.2     \\
Rotation                  & Rigid body                    \\ 
Initial angular momentum  & Kawaler (\cite{kawaler})      \\
Opacities                 & IR93 and AF94                 \\
Equation of state         & R96 and M88                   \\
Chemistry                 & Y=0.27, Z=0.0175              \\
Initial X(D)              & 2$\times$10$^{-5}$            \\ \hline
\end{tabular}
\end{table}

\begin{figure*}[htb]
%\centering
\centering{
\includegraphics[width=8cm]{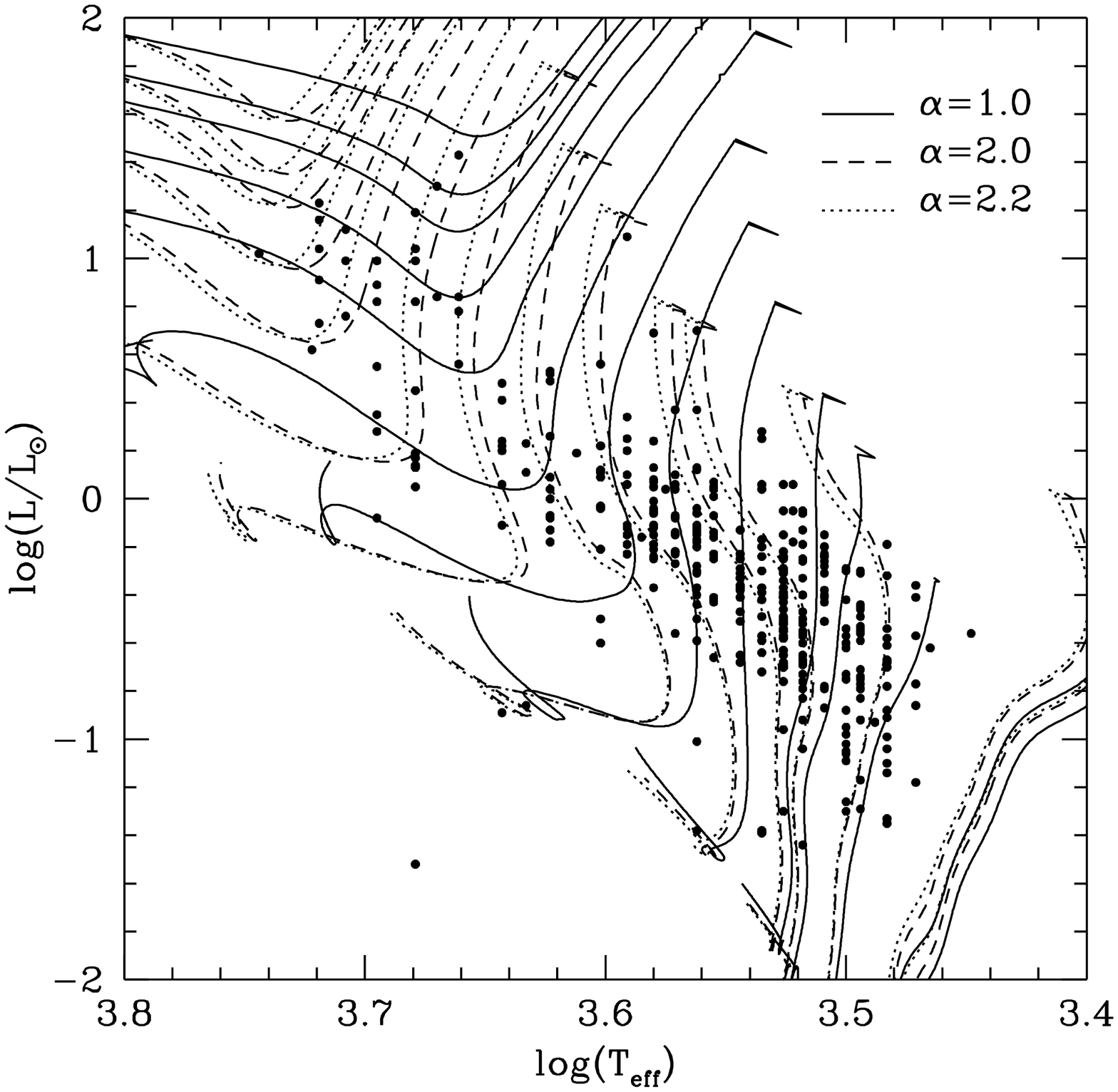}
\includegraphics[width=8cm]{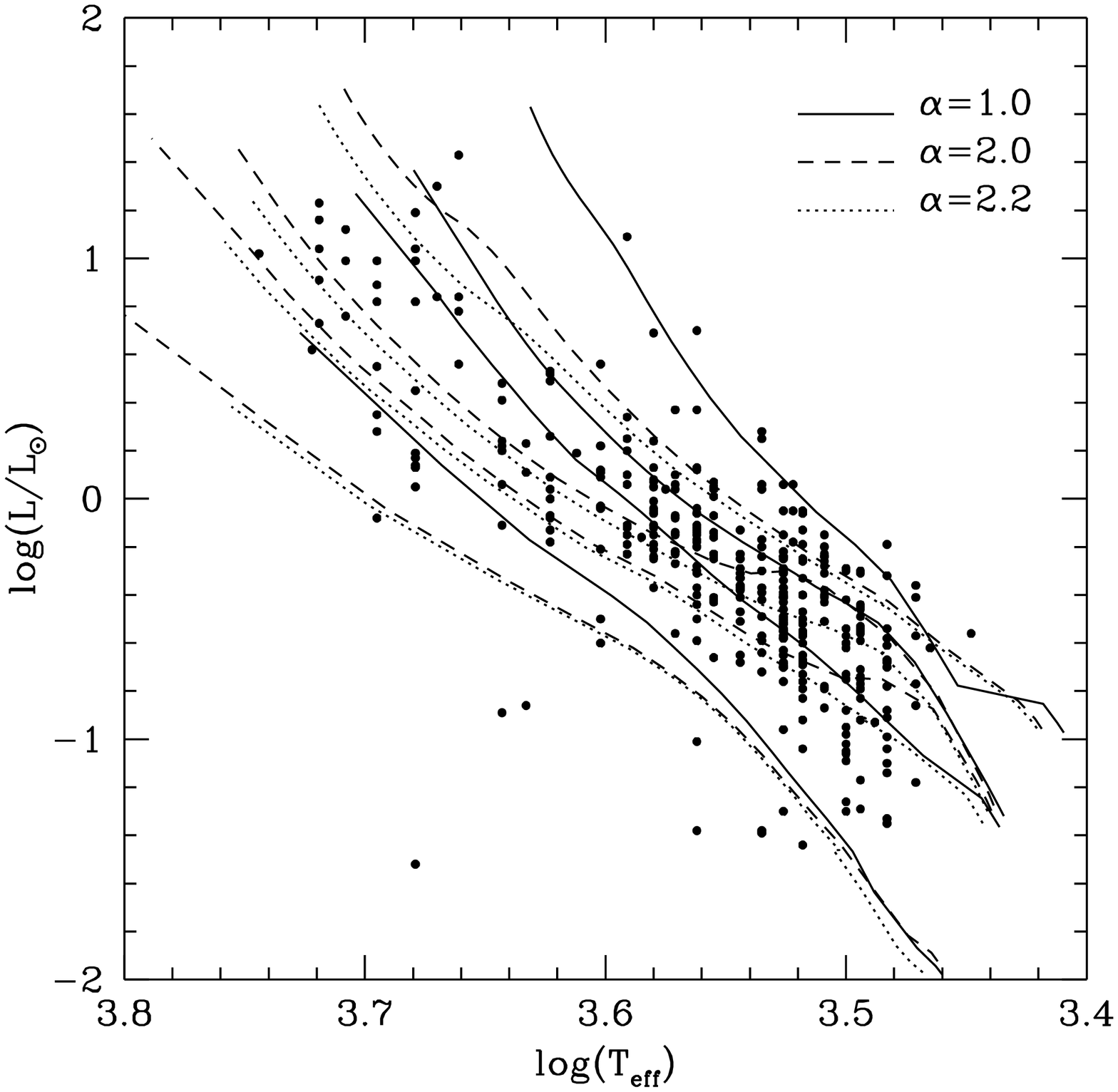}
}
\caption{Paths followed by the theoretical tracks (left) and isochrones 
(right) of the rotating
models calculated with three different values of the free parameter
$\alpha$ determining the mixing length. For clarity,
we report only the following masses calculated with
our three sets of tracks ($\alpha1.0$,$\alpha2.0$,$\alpha2.2$): 0.09, 0.1, 0.2, 0.3, 0.35, 0.5, 0.7, 1.0,
1.4, 2.0, 2.5, 3.0, 3.3 and 3.8~$M_{\odot}$ from bottom to top.
The observational data from Hillenbrand (\cite{hillen97}) are
represented by $\bullet$ symbols.
}
         \label{tracks}%
\end{figure*}
We computed pre-MS stellar evolutionary tracks in the mass range 
0.085$\leq$$M/M_{\odot}$$\leq$3.8. We adopted the solar chemistry with $Z$=0.0175 and 
$Y$=0.27, while the starting deuterium abundance in mass fraction, following 
Linsky (\cite{linsky}), was set to $X(D)$=2$\times$$10^{-5}$. 
The main physical parameters of the model sets are summarized in 
Table \ref{modpar}.
The evolutions start from a fully convective configuration with central 
temperatures in the range 5.3$<$$\log T_c$$<$5.8, follows deuterium and lithium 
burning and ends at the main sequence configuration.

\subsection{Boundary conditions}
We compute gray models for comparison with previous results. For the non-gray 
models we use the AH+AHS grid. Following a suggestion by Heiter et al.\ 
(\cite{heiter}), we match the atmospheric grid with the interior integration at 
$\tau_{\rm ph}$=10. This choice should minimize the consistency problems related to 
the different EOS and opacities adopted in the interior and in the atmosphere, 
and to the absence of turbulence pressure in the atmospheric modeling. 

\subsection{Choices about convection} \label{choiceconv}
As the FST atmospheric tables by Heiter et al.\ (\cite{heiter}) are available 
only for temperatures above $T_{\rm eff}$=4000~K, they cannot be used to 
compute models below $\sim 0.7${\msun} (Montalb\'an et al. \cite{josefina}). 
In particular, 
they cannot be used to analyze the ONC stellar population, mainly concentrated 
at $T_{\rm eff}$$<$4000~K. To achieve consistency between the internal and the 
atmospheric convective treatment, we decided to calculate only MLT models.

Apart from the quoted cases in which  $\alpha$ can be calibrated on hydrodynamic 
models, the necessity to simplify the numerical treatment of convection 
leads stellar modelists to use a single $\alpha$ for the convection zone and 
for all the evolutionary phases. This choice is equivalent to adopting an average 
efficiency of the convective transport on the whole extension of the convective 
region, and on all the evolutionary phases.
There is no good reason to assume that the $\alpha$ that, e.g., fits 
the solar radius should be used for other masses and for different evolutionary 
phases. Further, the efficiency of convection might change considerably within 
a convective zone, thus requiring the use of a variable $\alpha$. Thus a 
preliminary investigation of the effects of changing this parameter is 
mandatory.

We computed sets of models with three different values of the parameter: 
the models with $\alpha_{\rm in}$=2.0 ($\alpha2.0$ set) allow a fit of the solar
radius for non-rotating models\footnote{Also Baraffe et al.\ 
(\cite{baraffe}) and Montalb\'an et al.\ (\cite{josefina}) find a similar 
$\alpha_{\rm in}$ (=1.9) to reproduce the solar radius.}; the models with 
$\alpha_{\rm in}$=2.2 ($\alpha2.2$ set) are chosen to provide a ``very efficient convection" set. 
Both these sets are termed HCE (high efficiency convection) sets. 
We further provide tracks with $\alpha_{\rm in}$=1.0 ($\alpha1.0$ set). This latter choice, 
according to D'Antona \& Montalb\'an (\cite{franca2}, confirmed in \S \ref{lithium}), 
leads to a better agreement with the 
lithium vs. $T_{\rm eff}$ relation observed in young open clusters stars.
Remember, however, that the adopted BCs come from model atmospheres computed, 
down to $\tau_{\rm ph}$=10, with $\alpha_{\rm atm}$=1.0. 
(see the discussion in \S 2.5 and in Montalb\'an et al.\  
\cite{josefina}). We show in Fig.~\ref{tracks}\footnote{Fig.~\ref{tracks} and  
all others that use theoretical results were made with rotating models} 
the comparison between
tracks and isochrones of these three sets, showing the well known fact that both masses and ages
are affected by the choice of the convection model.

\subsection{Initial angular momentum}
Rotation was modeled according to the rigid body law. This choice is motivated by 
the fact that most of the low mass stars are still fully convective in 
the evolutionary 
stages of interest for this study. As a first attempt, the initial angular 
momentum $J_{\rm in}$ for all models was estimated according to the 
prescriptions by Kawaler (\cite{kawaler}). In that work, a relationship between 
angular momentum and stellar mass for stars earlier than F0 was derived using 
main-sequence (MS) stellar models and an estimate of the mean initial angular 
momentum-mass relation was made for stars of later spectral type. As  masses 
larger than about 1.5$M_{\odot}$ do not lose much angular momentum during 
their 
early evolutionary phases, it can be assumed that these stars reach the MS with 
the same angular momentum that they had at the beginning of their evolution. 
Kawaler (\cite{kawaler}) is able to reproduce the observational relation at
M$>$1.5M$_{\odot}$ using his own models for radiative stars, he then extends the
models to lower mass stars. For these, however, the observations
do not provide a direct comparison with the initial angular momentum, as
their rotation has been slowed down during the main sequence lifetime (but
not significantly during the pre--MS, if we adopt the hypothesis of pre--MS
disk locking in the same way as Bouvier et al.\ \cite{bouvier1997}). 
For the range 0.6\,-\,1.25M$_{\odot}$, the initial angular momentum-mass 
relation can be easily obtained from the respective mass-radius and 
mass-moment of inertia relations from Kawaler (\cite{kawaler}):

\begin{equation}
\centering
J_{\rm kaw}=1.566 \times 10^{50} \left( {M \over M_{\odot}} \right) ^{0.985}
~~~\mathrm{cgs}.
\label{kaweq}
\end{equation}

Once started with this initial angular momentum, we keep it constant
in our models during the pre--MS according to the above mentioned hypothesis,
whose validity will be checked later. 
This expression will then be extended to smaller masses as a result 
of the present study. 
The comparisons made in this work will help us to calibrate different models of 
angular momentum evolution.

\section{First comparisons among models}
\label{sect5}
\subsection{The role of the D-burning}
An uncertainty that deserves
a separate discussion is the abundance originally assigned to
deuterium. Deuterium burning, at least in the classic scenario,
takes place in the early phases of the pre-MS evolution, when
the central temperatures reach $T$$\sim$10$^6$K. A higher
deuterium abundance does not alter the track location,
but stretches the duration of the D-burning 
phase itself, thus determining a delay in the evolution.
We calculated for comparison a set of tracks with double deuterium abundance,
 $X(D)$$\sim$4$\times$$10^{-5}$, that is the highest value
acceptable on observational grounds (Linsky \cite{linsky}). 
In comparing the two sets of tracks, we note the following:
{\vspace{-.5\baselineskip}
\begin{enumerate}
\item{the track location does not change;}
\item{for $M$$>$0.5$\,M_{\odot}$, the high-D evolutions are slower,
but the age difference is in all cases shorter than 200,000yr;}
\item{for $M$$<$0.4$\,M_{\odot}$, the age differences between
tracks of the same mass are no longer negligible, 
from $\sim$250,000yr for $M$=0.4$\, M_{\odot}$ to 
$\sim$600,000yr for $M$=0.2$\,M_{\odot}$.}
\end{enumerate}
}
\noindent
In terms of the general conclusions found during the analysis
of the ONC stellar population, increasing the D abundance would 
have no effect on the mass distribution, while the age
distribution would be shifted to slightly older ages.

\begin{figure*}[htb]
\centering{
\includegraphics[width=8cm]{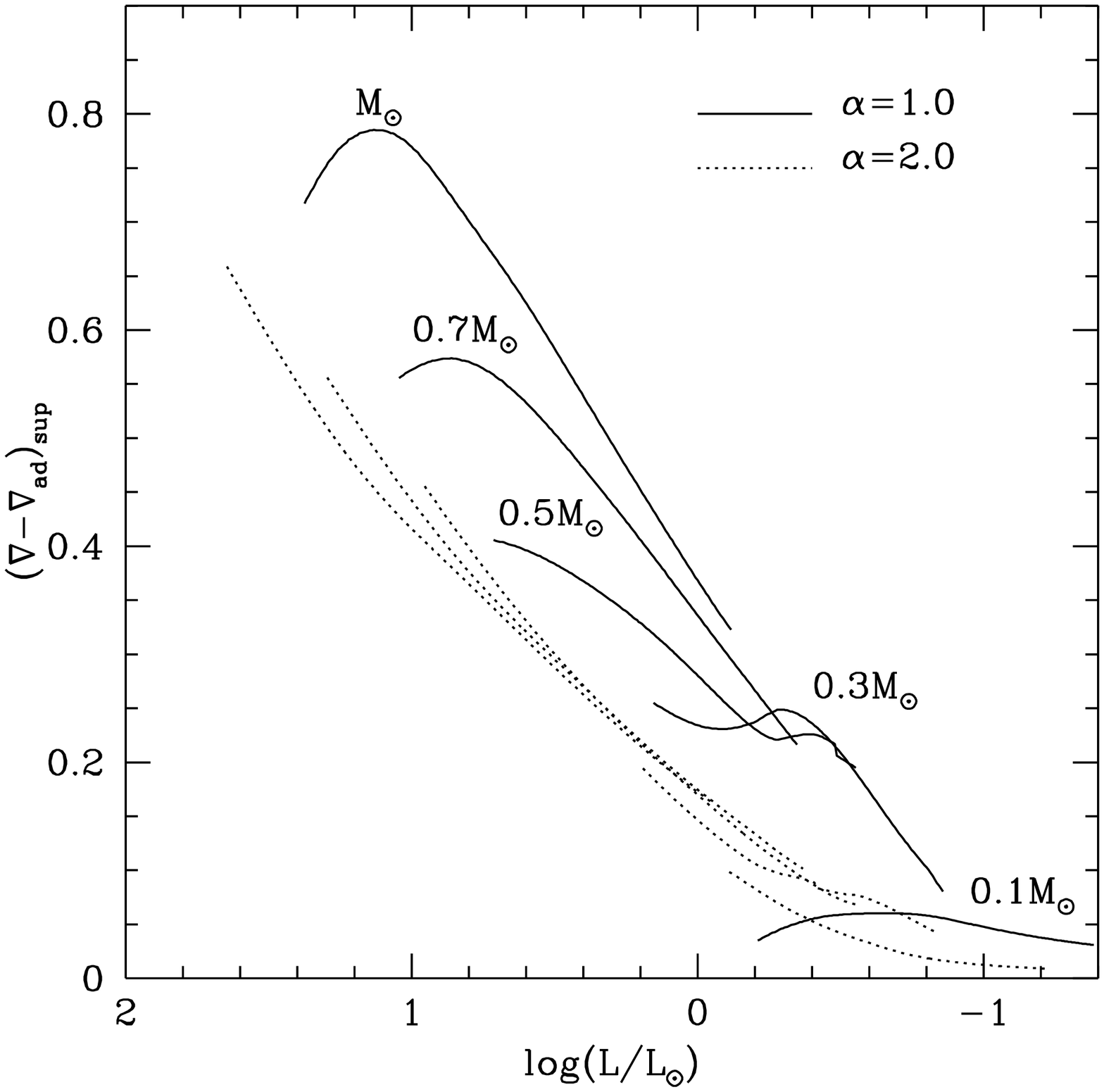}
\includegraphics[width=8cm]{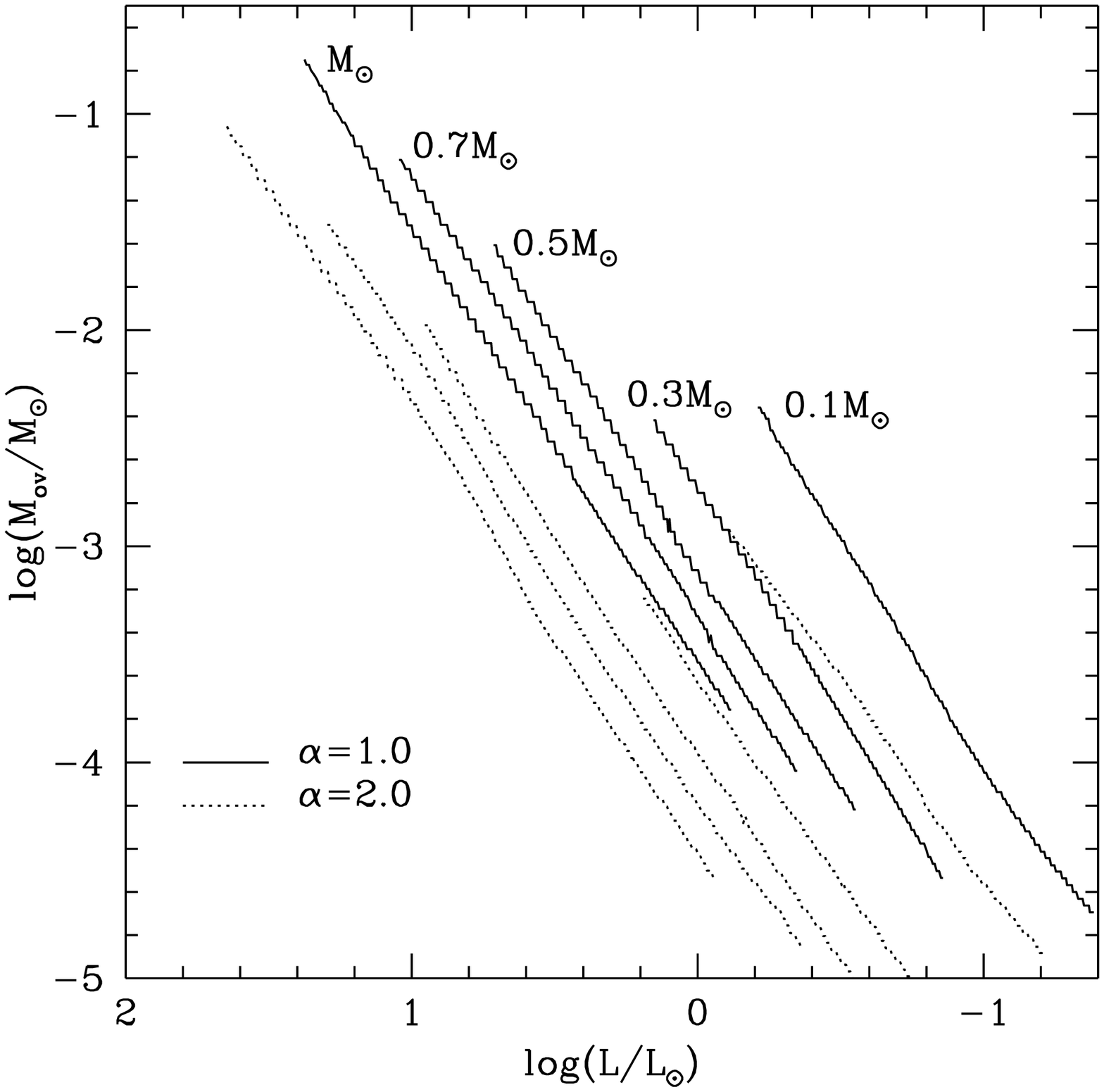}
           }
\caption{{\bf Left}: Variation with luminosity of the 
overadiabaticity of the stellar layer at $\tau$=10 of 
some pre-MS models calculated with
different values of the parameter $\alpha$ entering
the mixing length expression. {\bf Right}: The variation with
luminosity of the fraction of the mass of the star where 
$(\nabla-\nabla_{\rm ad})>10^{-4}$, for the same masses
shown in the left panel. For clarity we show only models $\alpha1.0$ and 
$\alpha2.0$.}
   \label{ovsup}
\end{figure*}

\subsection{Role of non-grayness} \label{gray}
The gray tracks of models with $\alpha$=1.0 are systematically hotter 
than their non-gray counterparts, although the difference in 
$T_{\rm eff}$ varies with the mass. For the lowest masses of our sample,
i.e.\ $M$$\leq$0.2$M_{\odot}$, the high gravities allow the differences  
to remain within $\Delta T_{\rm eff}$$\sim$100$\,$K. This difference increases
to $\sim$250$\,$K at 0.6$M_{\odot}$ and reaches a maximum of $\sim$400$\,$K in 
1$M_{\odot}$ models only at the end of the Hayashi track.
The stars for which the pre-MS tracks are most sensitive to the 
boundary conditions are those with masses in the range
$0.4M_{\odot} \leq M \leq 1M_{\odot}$: the difference in $T_{\rm eff}$ 
slightly increases along each track, and reaches a maximum of
$\Delta T_{\rm eff} \sim 400$K when the radiative core is formed.  
The $T_{\rm eff}$ of more massive stars are less influenced by the
atmospheric treatment.
These differences in $T_{\rm eff}$ lead to the assignment of a different mass to 
a given observed star. We verify that, consistent with the above discussion, 
this effect is negligible for the lowest masses, while it leads to differences 
of the order of $\sim$0.1$M_{\odot}$ for $M$$\sim$0.3$-$0.5$M_{\odot}$, where, 
as we shall see, the bulk of the ONC population is found. For larger masses the 
differences are larger because the tracks for different masses are closer to each 
other.

\subsection{The convection model} \label{convec}

The tracks corresponding to the three values of $\alpha$ used in the present 
investigation are shown in the left panel of the Fig.~\ref{tracks}. The 
observed stellar loci are also reported. In the early phases, with the 
exception of the D-burning phase, gravitational contraction is the only source 
of energy. The densities and temperatures increase until 
the central regions become radiatively stable; shortly after, H-burning takes 
over as the main energy source. We see from the left panel of the 
Fig.~\ref{tracks} that the $\alpha2.0$ and $\alpha2.2$ tracks are 
systematically hotter, the differences being larger for higher masses and 
smaller for older ages. This behavior can be understood on the basis of the 
different degree of overadiabaticity present in the various masses at different 
ages.  
The exact temperature profile depends on the overadiabaticity 
($\nabla - \nabla_{\rm ad}$), which is defined as the excess of the effective 
temperature gradient in comparison with the adiabatic temperature gradient, 
and it is only noticeably 
different from zero at the border of the convective zone, where convection
becomes inefficient. 
The two panels of Fig.~\ref{ovsup} 
show, respectively, the evolution of 
the overadiabaticity at $\tau$=10 and of the width (in solar masses) 
of the external 
region of the star where  $(\nabla - \nabla_{\rm ad}) > 10^{-4}$. For clarity 
we report only the $\alpha1.0$ and $\alpha2.0$  
models. The luminosity is on the abscissa as a time indicator. A detailed 
inspection of Fig.~\ref{ovsup} shows:
\begin{enumerate}
\item{the overadiabaticity at $\tau$=10 of the $\alpha1.0$ models is systematically
higher than their $\alpha2.0$ counterparts. This can be understood on the 
basis of the intrinsic lower efficiency of the convective model adopted,
due to a lower mixing length;}
\item{due to the higher internal densities of the less massive models (hence, 
greater convective efficiencies), the 
overadiabaticity differences increase with the mass;}
\item{in the less massive models the differences above tend to narrow with age;}
\item{the extension in mass of the overadiabatic region is also systematically 
higher in the $\alpha1.0$ models, and tends to shrink with age.}
\end{enumerate}
We verified that models of the same mass belonging to the two sets of tracks
follow the same $L(t)$ relation. The differences in the location of the tracks
are, therefore, to be totally ascribed to differences in the
effective temperatures.
Since the interior of these structures is practically adiabatic 
(in the center, $(\nabla -\nabla_{\rm ad})$$<$$10^{-7}$ in all cases), the radius is 
mainly determined by the degree of the overadiabaticity. 
This explains why larger differences are found for higher masses. 
In the lowest masses the efficiency of convection increases at 
older ages, so that the sensitivity to the 
adopted model for convection is strongly reduced. The low mass tracks 
approach each other at low luminosities (see Fig.~\ref{tracks}).

\begin{figure*}[htb]
\centering{
\includegraphics[width=8cm]{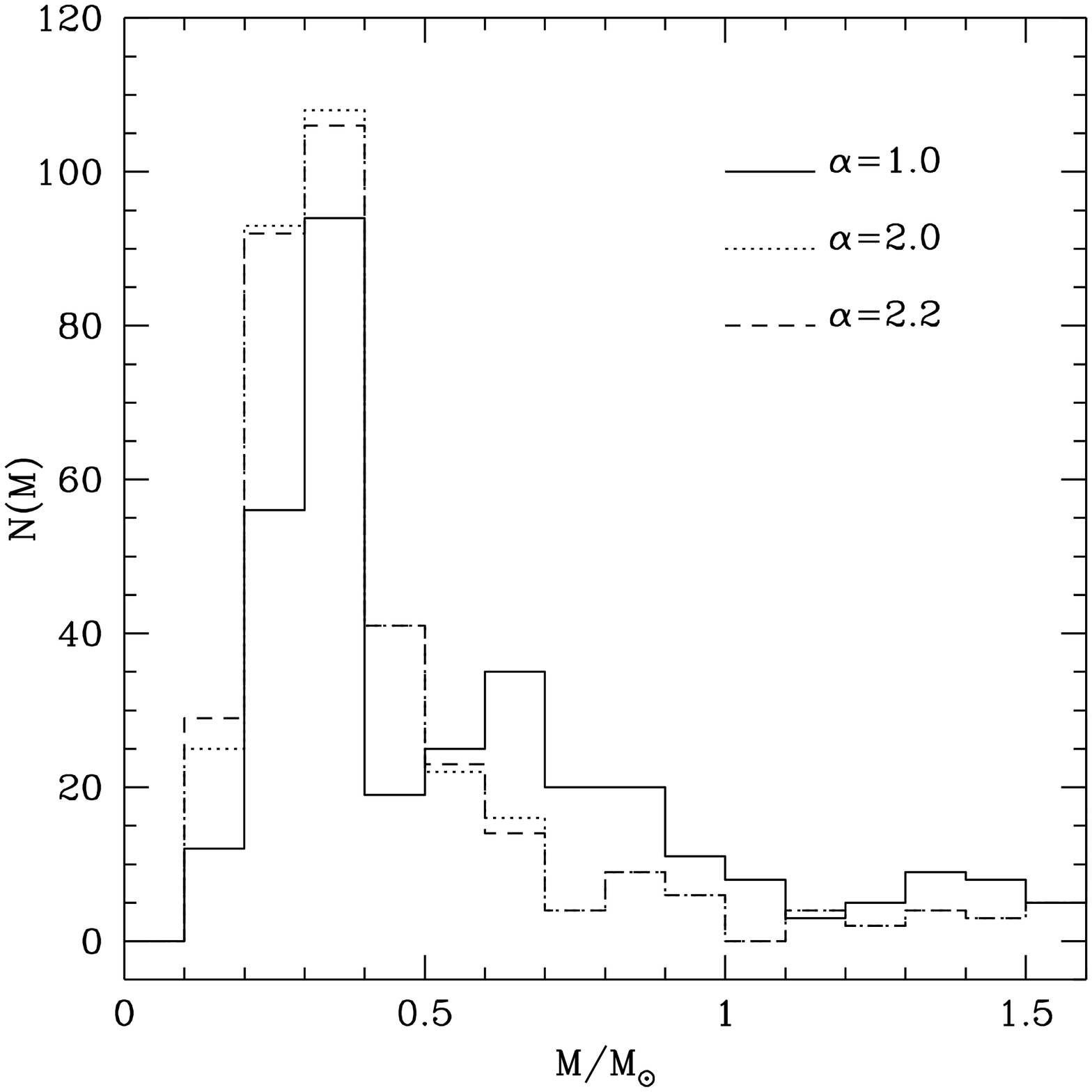}
\includegraphics[width=8cm]{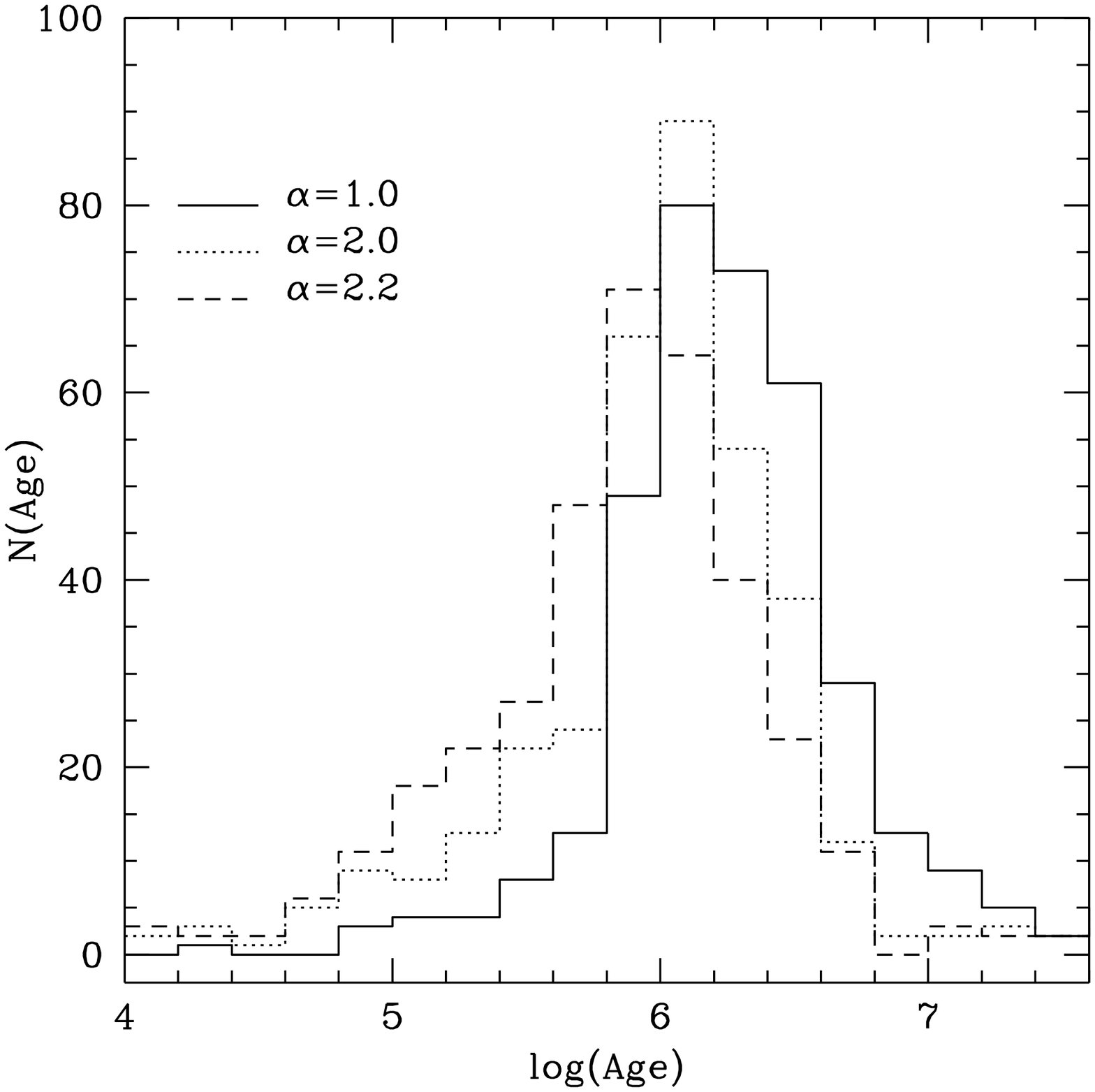}
           }
\caption{Mass (left) and age (right) histogram based, respectively, on the 
masses and ages assigned to the observed stars in the ONC using three 
different sets of tracks, calculated with three prescriptions for convection,
$\alpha$=1.0, 2.0 and 2.2.
}
   \label{age}
\end{figure*}
\subsection{The lithium depletion}\label{lithium}

\begin{table}[b]
\caption{Lithium abundances at the age of $10^8$yr for 0.7, 0.8, 0.9, 1.0 and
1.2 M$_{\odot}$. We report the values obtained with different sets of our 
non-gray models, with (``rot'') and without (``non-rot'') rotation. The initial lithium concentration is $\rm{\log N[Li]=3.31}$.}
\label{table-lithium}
\centering
\setlength{\tabcolsep}{0.6\tabcolsep}
\begin{tabular}{lccccc}
\hline \hline
\multicolumn{1}{c}{Models} & 0.7M$_{\odot}$ & 0.8M$_{\odot}$ & 0.9M$_{\odot}$ & 1.0M$_{\odot}$ & 1.2M$_{\odot}$  \\ \hline
rot $\alpha$=1.0 & $\hphantom{-}$0.750 & $\hphantom{-}$2.607 & $\hphantom{-}$3.087 & $\hphantom{-}$3.226 & $\hphantom{-}$3.291      \\ [-2pt]
rot $\alpha$=2.0 & $-$0.781 & $\hphantom{-}$1.669 & $\hphantom{-}$2.267 & $\hphantom{-}$2.720 & $\hphantom{-}$3.139     \\ [-2pt]
rot $\alpha$=2.2 & $-$0.961 & $\hphantom{-}$1.466 & $\hphantom{-}$2.109 & $\hphantom{-}$2.592 & $\hphantom{-}$3.094     \\ [-2pt]
non-rot $\alpha$=1.0 & $\hphantom{-}$0.950 & $\hphantom{-}$2.650 & $\hphantom{-}$3.097 & $\hphantom{-}$3.228 & $\hphantom{-}$3.291  \\ [-2pt]
non-rot $\alpha$=2.0 & $-$0.524 & $\hphantom{-}$1.715 & $\hphantom{-}$2.295 & $\hphantom{-}$2.733 & $\hphantom{-}$3.143 \\ [-2pt]
non-rot $\alpha$=2.2 & $-$0.521 & $\hphantom{-}$1.712 & $\hphantom{-}$2.297 & $\hphantom{-}$2.733 & $\hphantom{-}$3.143 \\
\hline
\end{tabular}
\end{table}

Table \ref{table-lithium} shows the lithium concentrations for our three 
sets of tracks and their non-rotating counterparts. We report the values for
0.7, 0.8, 0.9, 1.0 and 1.2 M$_{\odot}$ at $10^8$yr. A comparison between
$\rm{\log N[Li]}$ found with rotating and non-rotating models, keeping $\alpha$
fixed, is qualitatively in agreement with the abundances found by 
Mendes et al.\ (\cite{mendes}), i.e., rotating models provide greater lithium
depletion especially for low-mass stars at the age in question. Another 
comparison between the abundances found with different values of $\alpha$,
keeping the rotation status fixed,
confirms the results by D'Antona \& Montalb\'an (\cite{franca2}): the lithium
depletion of the HCE models is too large to be consistent with the lithium 
depletion observed by Soderblom et al.\ (\cite{soder93}) and Garcia Lopez
et al.\ (\cite{garcia94}) in young open clusters, which can be reproduced only by 
the LCE $\alpha$1.0 models. On the contrary, the solar radius is reproduced only by
the $\alpha$2.0 model, and 2D hydrodynamic computations indicate HCE in the
pre-MS. We regard this result as an indication that the efficiency of convection
in the pre-MS might be affected by other parameter(s).
The "second parameter" affecting lithium depletion is identified as the
stellar rotation rate by Siess \& Livio (\cite{siess}) and in the papers
by Ventura et al. (\cite{ventura}) and D'Antona et al. (\cite{dantonaetal2000}). 
The first authors
propose that $\alpha$ is smaller in fast rotating pre--MS stars due
to the twisting of convective cells, the others show that the action
of the dynamo-induced magnetic field due to the interaction of
rotation and convection modifies the structure of the convective layers
and reduces lithium depletion.

\section{Comparison with the ONC stars}
\label{sect6}

\subsection{Data from the literature}
In order to study angular momentum evolution in pre-MS phase, we will compare 
our sets of evolutionary tracks with observational data of the ONC stars. To 
accomplish this goal we need some key parameters, such as effective 
temperatures and luminosities (to infer masses and ages of the stars) and 
also the rotation period and an index that allows us to distinguish between 
different kinds of angular momentum evolutions. 

\begin{figure*}[htb]
\centering{
\includegraphics[width=5.9cm]{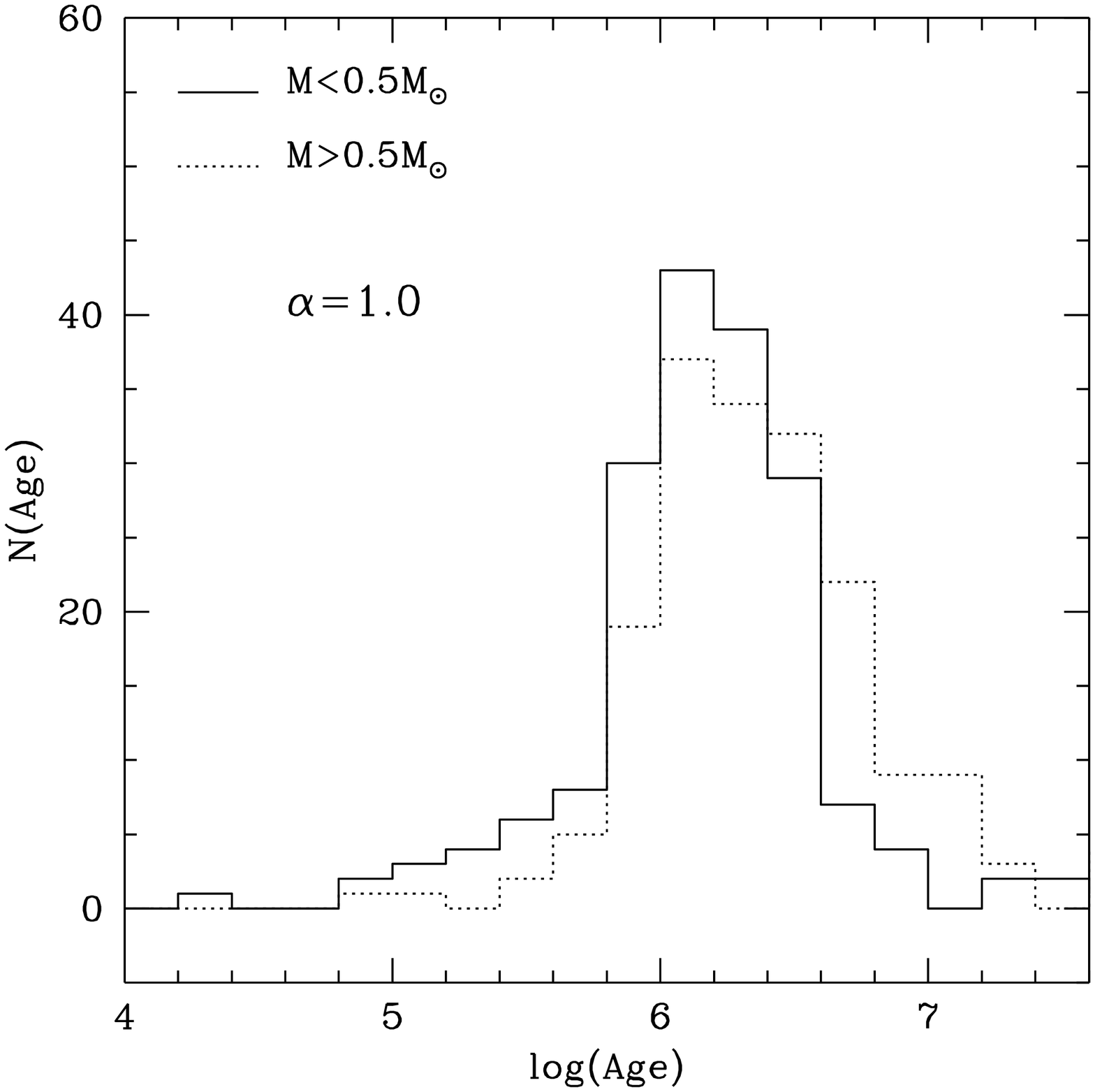}
\includegraphics[width=5.9cm]{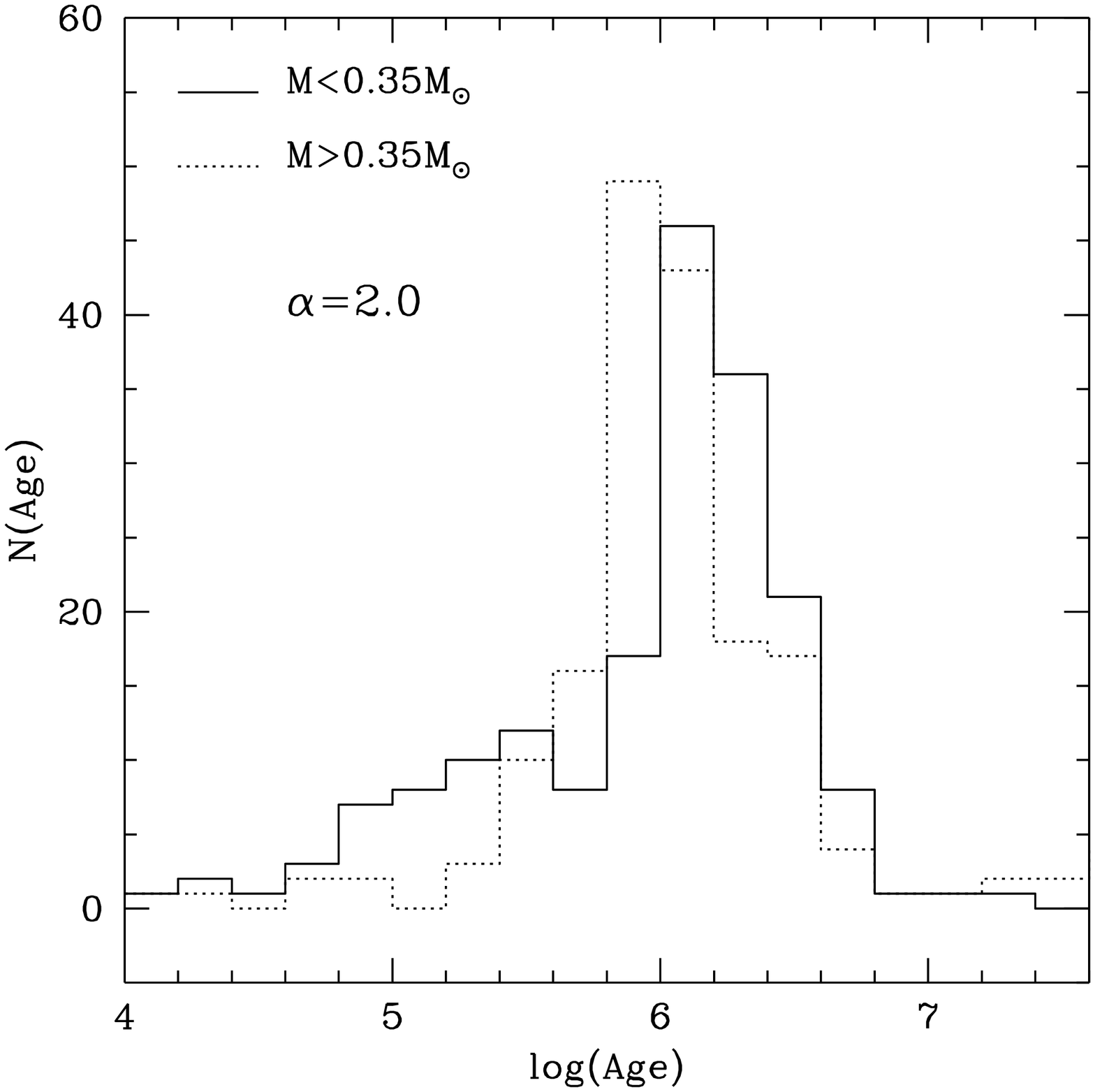}
\includegraphics[width=5.9cm]{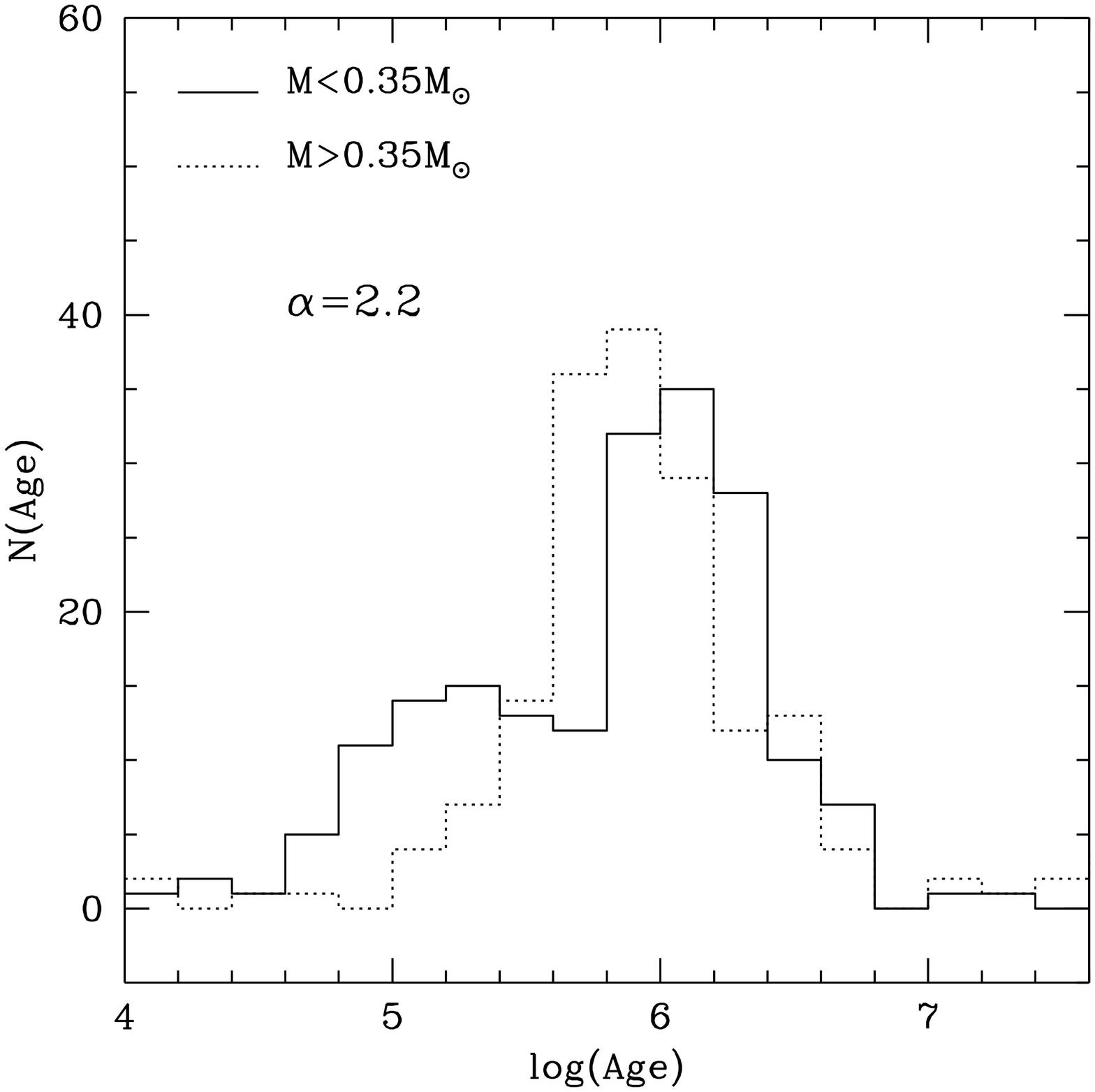}
           }
\caption{The comparison between the age distributions
of the observed stars with assigned mass lower and higher than M$_{\rm tr}$. The 
M$_{\rm tr}$ value is 0.5M$_{\odot}$ for LCE models and 0.35M$_{\odot}$ for HCE
models.}
   \label{agemass}
\end{figure*}

The ONC data we used have been kindly provided by Dr.\ Keivan Stassun, who has widely
worked on the rotational properties of ONC (Stassun et al.\ 
\cite{stassun}, \cite{stassun04}). Our final study sample is composed 
of a combination of data from the following sources:
\begin{enumerate}
\item{Rotation periods: Stassun et al.\ (\cite{stassun}), Herbst
et al.\ (\cite{herbst02});}
\item{Effective temperatures and luminosity:
Hillenbrand (\cite{hillen97});}
\item{Infrared continuum excess, $\Delta$[I$-$K]: Hillenbrand 
et al.\ (\cite{hillen98});}
\item{\ion{Ca}{II} equivalent width, EW(\ion{Ca}{II}): Hillenbrand (\cite{hillen97}),
Hillenbrand et al.\ (\cite{hillen98});}
\item{X-ray luminosities: Stassun et al.\ (\cite{stassun04}).}
\end{enumerate} 
The rotation period diagnostic was obtained by photometry, interpreting as 
rotation the periodic photometric variability, probably due to the presence of 
stellar spots.

The IR excess ($\Delta[$I$-$K]) was obtained from combined
optical and infrared photometric data. Extinction and spectral typing uncertainties
are the main sources of errors in $\Delta$[I$-$K] estimates. For earlier
spectral type stars ($\lesssim$K2) the errors are negligible
($<$0.05 mag), for spectral types in the range K2$-$M3, typical errors are between
0.1$-$0.3 mag, and it is largest for the latest spectral types, where
mis-classification causes relatively larger errors. 

The EW(\ion{Ca}{II}) was obtained from the optical spectroscopic study of
Hillenbrand (\cite{hillen97}) and analyzed by 
Hillenbrand et al.\ (\cite{hillen98}). Their measurements uncertainty is
estimated at 0.5{\AA} based on measurements of multiple spectra
of the same star. 

In order to obtain more reliable values for $L_{\rm X}$ of ONC stars, 
Stassun (\cite{stassun04}) reanalyzed all archival {\it Chandra}/ACIS 
observations of these objects using updated calibrations and including 
time-filtering of flares.

The effective temperatures and luminosities were obtained from optical
spectroscopy and photometry. 

\subsection{Derivation of masses and ages}\label{magedev}

For each of the three sets of tracks previously discussed, we assigned to each 
observed point a mass and an age by linearly interpolating between the two 
nearest tracks. An inspection of Fig.~\ref{tracks} can help understand, at 
least qualitatively, the differences that we should find by varying $\alpha$. 
For each observed star in the $\alpha2.0$ and $\alpha2.2$ sets, we assign a 
systematically smaller mass than in the $\alpha1.0$ models, 
hence a younger age (we recall that this 
evolutionary phase is governed by gravitational contraction, that proceeds on a 
Kelvin-Helmoltz time scale $\tau_{\rm KH}$$\propto$${M^2 \over RL}$). If the 
$\alpha2.0$ or $\alpha2.2$ sets are used, we therefore expect a mass 
distribution shifted to lower masses and, on the average, a younger population. 
Typical internal errors are $\lesssim$0.2 dex in $\log (L/L_{\odot})$ for all 
spectral types and are $\lesssim$0.02 dex in $\log T_{\rm eff}$ for late-type 
(K$-$M) stars, but increase towards earlier spectral types. This leads to an 
uncertainty in the determination of mass that is $\lesssim$0.1$M_{\odot}$ for 
$M$$<$0.5$M_{\odot}$, and gradually increases to $\sim$0.2$M_{\odot}$ for 
$M$$\sim$$1M_{\odot}$. The attribution of age is mainly influenced by the 
uncertainty on the luminosity, that makes the age uncertain by $\sim$1$\,$Myr 
at the age of 1$\,$Myr. This poses the problem of whether the age distribution 
we find should be considered either as the result of a burst of star formation 
or as a real 
indication of age differences from star to star. In the course of the 
investigation we favour a statistical interpretation of data as an 
indication of some age evolution, based on the evolution of rotation periods. 

The left panel of the Fig.~\ref{age} shows the mass distribution of the 
observed stars, obtained by using the three sets. 
Only masses $M$$<$1.6$\,M_{\odot}$
are plotted because they represent most of the 
stars in the sample. The mass function for the $\alpha1.0$ set peaks in the 
mass interval 0.3$-$0.4M$_{\odot}$, but we also note the presence of a 
significant group of stars with masses in the range 0.2$-$0.3M$_{\odot}$ and 
another group in the interval of 0.6$-$0.9$\,M_{\odot}$. For the HCE models this
latter group of objects becomes less relevant and the mass function peaks in 
the mass interval 0.2$-$0.4M$_{\odot}$.
The right panel of Fig.~\ref{age} confirms that the age distribution depends on 
the choice of $\alpha$. We note that a slightly younger population is obtained 
as the value of $\alpha$ increases, and, in the $\alpha$2.0 and $\alpha$2.2 
cases, a 
very young group of stars appears at ages $\sim$1$-$2$\times$$10^5$yr,  but is not 
present for the LCE ($\alpha1.0$) models. This can be understood  by 
considering, in the right panel of Fig.~\ref{tracks}, the relative location of 
observed points and theoretical isochrones. 
The age differences are also due to the different slope of the isochrones 
corresponding to the two sets of tracks, which, in turn, are related to the 
differential variation of temperature with mass, already outlined in 
\S\ref{convec}.
In the HCE models there is a group of objects with ages clustering around 
$100,000$ yr . For the $\alpha$2.0 case this young group is made up of $\sim$40 
stars and for the $\alpha2.2$ set its presence is still more evident, 
exhibiting $\sim$70 objects. 

In \S \ref{sec-dichotomy} we will define a transition mass M$_{\rm tr}$ for
HCE and LCE models, on the basis of the rotational periods distribution. Here
we compare, in Fig.~\ref{agemass}, the age distribution for two different 
ranges of mass, M$>$M$_{\rm tr}$ and M$<$M$_{\rm tr}$. We see from the 
left panel of Fig.~\ref{agemass} that in the $\alpha1.0$ set the two 
populations show a similar distribution. 
On the contrary, the age
distribution of the two groups for HCE models is very different.
%: while
%the bulk of the population of the less massive stars is on the average older,
%we see that the very young population shown in Fig.~\ref{age} 
%is present only in the low masses group, and this is due to the shape of 
%the low mass tracks in the HCE sets (see Fig.~\ref{tracks}).
Thus, while the existence of a group of younger stars would be possible
in the formation history of the ONC, it should be present for any mass.
The discrepancy in the age distribution may be again an indication 
that the $\alpha1.0$ set provides a 
better description of the ONC stellar population,  in agreement with the 
lithium depletion discussed in \S  \ref{lithium} and with the previous analysis 
made by D'Antona \& Montalb\'an (\cite{franca2}). In any case, this certainly 
is not final and we proceed with the analysis by using the 
three sets of tracks.

\begin{figure}[t]
\includegraphics[width=8cm]{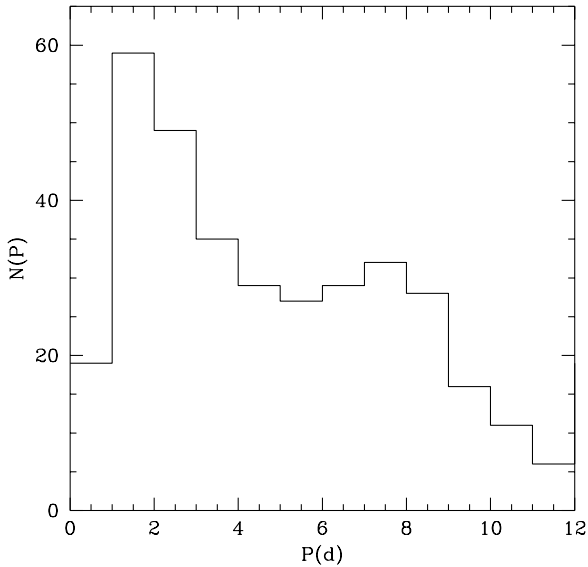}
\caption{The period histogram of all observed ONC objects. The total 
distribution of periods has a bimodal character, showing a primary peak of
fast rotators (P$\sim$2 days) and a secondary peak of slow rotators 
(P$\sim$8 days).}
   \label{histop}
\end{figure}

\subsection{Comparison with gray models}\label{ngconv}

We compare now the effects introduced by the two major 
factors, i.e. convection and boundary conditions. Although for masses 
$M$$>$$1M_{\odot}$ the treatment of convection is more relevant than the boundary 
conditions adopted in determining the effective temperature of the tracks 
(Montalb\'an et al.\ \cite{josefina}), for the interesting range of mass for 
the ONC comparison, namely 0.2$-$0.4$\,M_{\odot}$, and the non-gray models, 
convection is important mainly in the early evolutionary phases. At later 
phases (ages $>$1Myr), convection becomes more adiabatic and the non-grayness 
becomes the main factor affecting the track location.  

For gray atmospheric treatment, keeping $\alpha$ fixed to 
$1.0$ would concentrate the same mass distribution at $M$$\sim$0.2$-
$0.3$\,M_{\odot}$. There would also be a considerable reduction of the 
population with masses in the range 0.6$<$$M/M_{\odot}$$<$1. A 
larger effect would be obtained if we had used gray models with $\alpha$=2.0. In 
this case, the mass function would be peaked in the range 
0.1$<$$M/M_{\odot}$$<$0.3, and the average age of the observed 
stars would be slightly younger than $\sim$1Myr. This latter result was 
obtained by Herbst et al.\ (\cite{herbst02}) during their analysis of the ONC 
population using the tracks by D'Antona \& Mazzitelli (\cite{franca1}) 
that use gray approximation and the very efficient FST model for convection. 

\section{Stellar rotation in the ONC} \label{poponc}

\begin{figure*}
\centering{
\includegraphics[width=5.9cm]{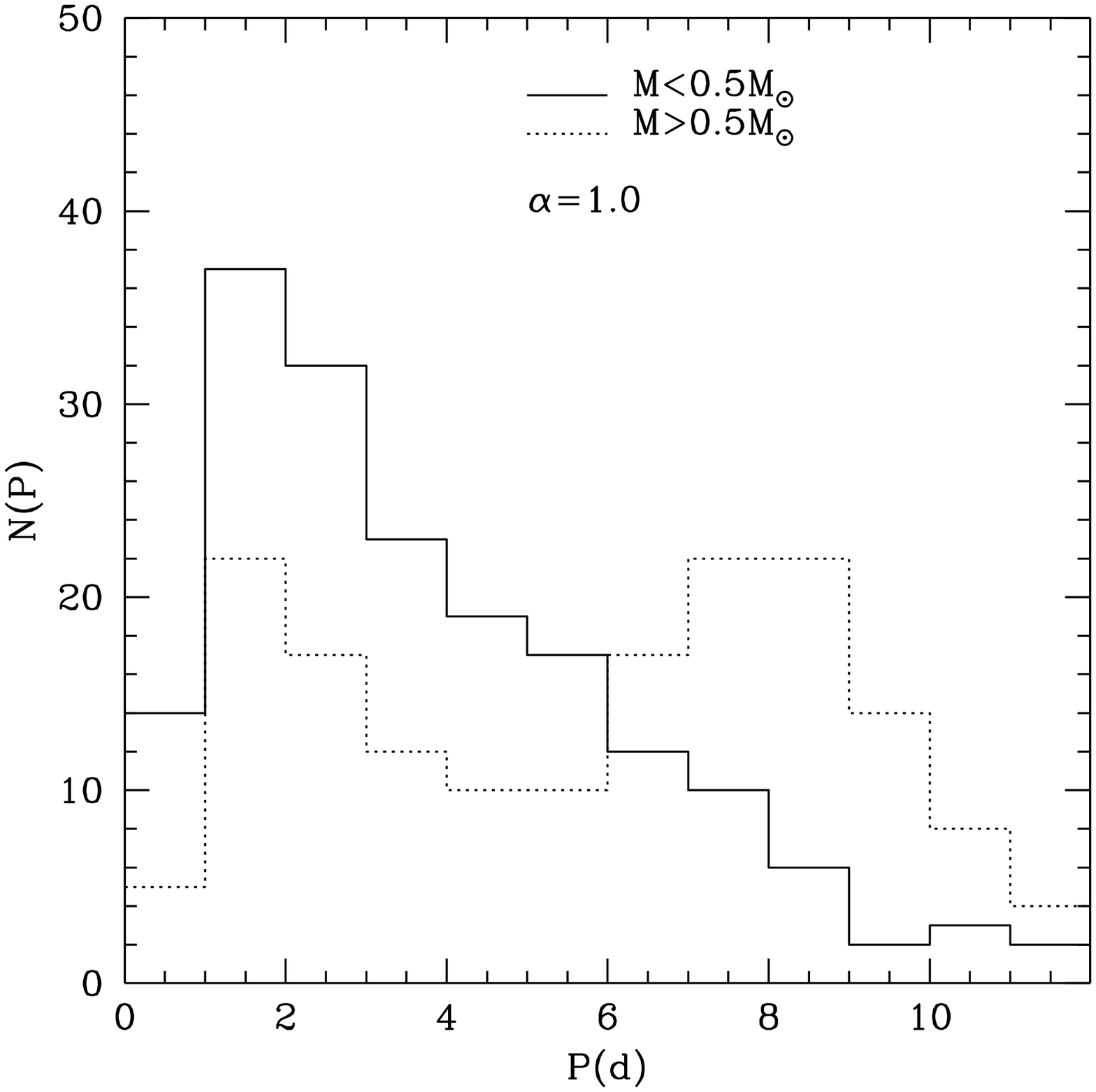}
\includegraphics[width=5.9cm]{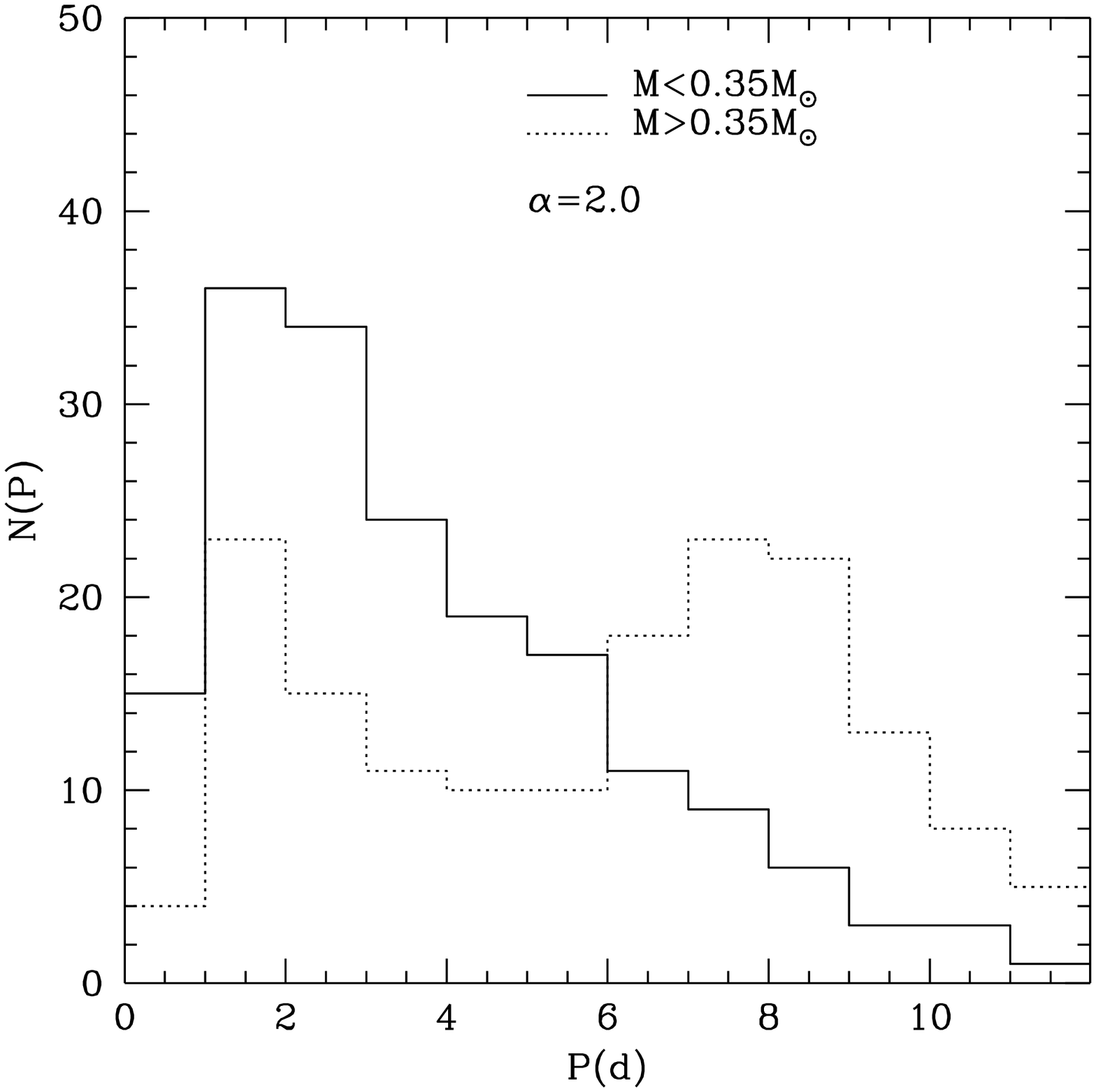}
\includegraphics[width=5.9cm]{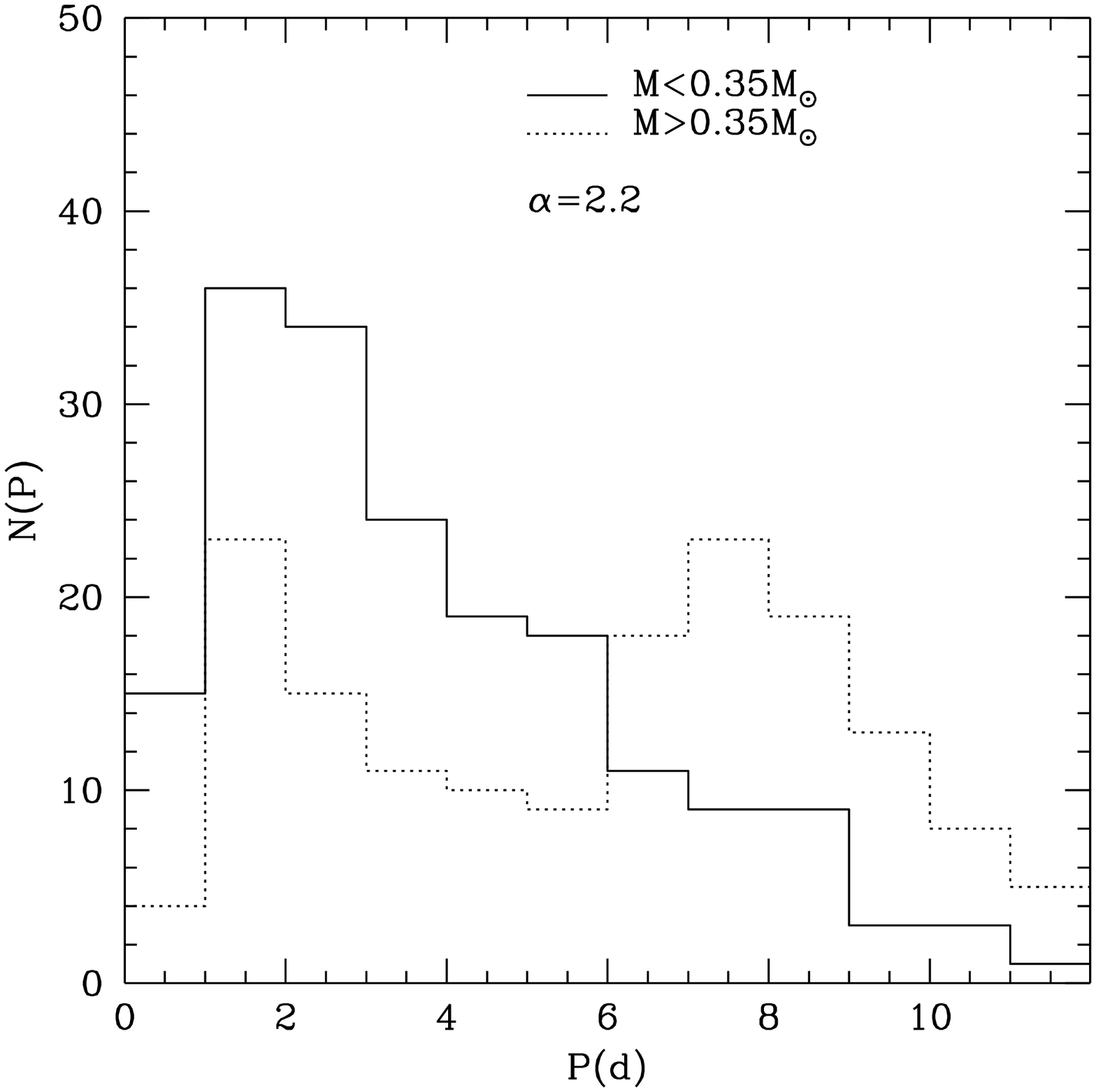}
           }
\caption{Period histograms showing the dependence 
on mass of the period distribution of the ONC objects. 
Stars more massive than M$_{\rm tr}$ have a
bimodal period distribution and their less
massive counterparts rotate faster and exhibit a unimodal distribution. 
This behavior can be seen independently of the choice of $\alpha$. 
The M$_{\rm tr}$ value is the same used in Fig.~\ref{agemass}.} 
   \label{dichotomy}
\end{figure*}

\subsection{The dichotomy in period distribution for
different mass ranges \label{sec-dichotomy}}
We examine  in Fig.~\ref{histop} the distribution of the observed rotation 
periods. We note the presence of a primary peak corresponding to fast rotators 
with 1$\lesssim$P(d)$\lesssim$3, and a secondary peak at P$\sim$8d. The former 
can be associated with spin up due to the conservation of the total 
angular momentum. The latter indicates the presence of a mechanism acting to 
prevent stellar spinning up, at least in the early evolutionary phases. 
Attridge \& Herbst (\cite{attridge}), Choi \& Herbst (\cite{choi}) suggested 
that this can be due to a ``disk locking'', due to magnetic coupling between 
the star and the disk (K\"onigl \cite{konigl}). Following Herbst et al.\ 
(\cite{herbst02}) we investigate in detail the rotational status of the 
various masses involved. 
The rotational properties of the stars vary considerably with mass:
stars with masses larger than a value M$_{\rm tr}$ have a clearly bimodal 
distribution, while the less massive sample contains only a tail of slow 
rotators. This behavior was first observed by 
Attridge \& Herbst (\cite{attridge}) and discussed by Herbst et al.\ 
(\cite{herbst02}). 
\begin{table}[t]
\caption{Main physical parameters of the present gray (G) and non-gray (NG)
models. $\cal{N}_<$ ($\cal{N}_>$) is the percentage of the N$_{\rm t}$ stars
that have mass less (greater) than M$_{\rm tr}$ for different rotation
periods ($P$). See text for
details.} \label{results}
\centering
\setlength{\tabcolsep}{0.6\tabcolsep}
\begin{tabular}{lcccccc}
\hline \hline
\\ [-9pt]
Models   &$\displaystyle{\rm Mass}\atop{\displaystyle\rm range}$ &$\displaystyle{\rm Age}\atop{\displaystyle\rm range}$ &M$_{\rm tr}$ &
$\cal{N}_<$&
$\cal{N}_<$&
$\cal{N}_>$ \\
 &(M$_{\odot}$)&(Myr)& (M$_{\odot}$) & $P$$<$4d &   $P$$>$6d  & $P$$>$6d  \\
\hline
G $\alpha$=1.0  & 0.2$-$0.3    & 0.6$-$2.5 &0.35 & 65\% & 19\% & 53\% \\ [-2pt]
G $\alpha$=1.5  & 0.1$-$0.3    & 0.3$-$1.3 &0.25 & 65\% & 19\%  & 54\% \\ [-2pt]
NG $\alpha$=1.0 & 0.2$-$0.4    & 1$-$2     &0.5  & 63\% & 18\% & 53\% \\ [-2pt]
NG $\alpha$=2.0 & 0.2$-$0.4    & 0.6$-$2.5 &0.35 & 67\% & 18\%  & 55\% \\ [-2pt]
NG $\alpha$=2.2 & 0.2$-$0.4    & 0.4$-$1.6 &0.35 & 67\% & 23\%  &54\% \\ \hline
\end{tabular}
\end{table}
We define the ``transition mass" (M$_{\rm tr}$), which depends 
on the track set chosen for the analysis, 
in order to maximize the effect of the bimodality.
Although the dichotomy does not depend on 
the chosen set, the transition mass varies according to it.
For LCE models it is 0.5M$_{\odot}$, while a reasonable value 
 is 0.35M$_{\odot}$ for HCE models. If we had
used our gray models with $\alpha$=1.5, M$_{\rm tr}$ would have been 
still smaller, namely, 0.25M$_{\odot}$, in agreement with the findings
by Herbst et al.\ (\cite{herbst02}),
who used the HCE FST models by D'Antona \& Mazzitelli (\cite{franca1}).
In Fig.~\ref{dichotomy} we
show the histogram of periods, respectively, for stars less and more massive
than M$_{\rm tr}$. The secondary peak at $P$$\sim\,$8d, already 
seen in the Fig.~\ref{histop}, is present only in the population
at M$>$M$_{\rm tr}$, while
the low mass objects show a clear trend towards short periods. 
Table \ref{results} shows the percentages of slow and fast rotators
(here defined by the limitations P$>$6d and P$<$4d).
Fast rotators contains more than 60\% of masses $<$M$_{\rm tr}$. 
On average, only 20\% of
the masses $<$M$_{\rm tr}$ and $\sim$54\% of the masses $>$M$_{\rm tr}$ have P$>$6d.

\begin{figure*}[htb]
\centering {
\includegraphics[width=5.9cm]{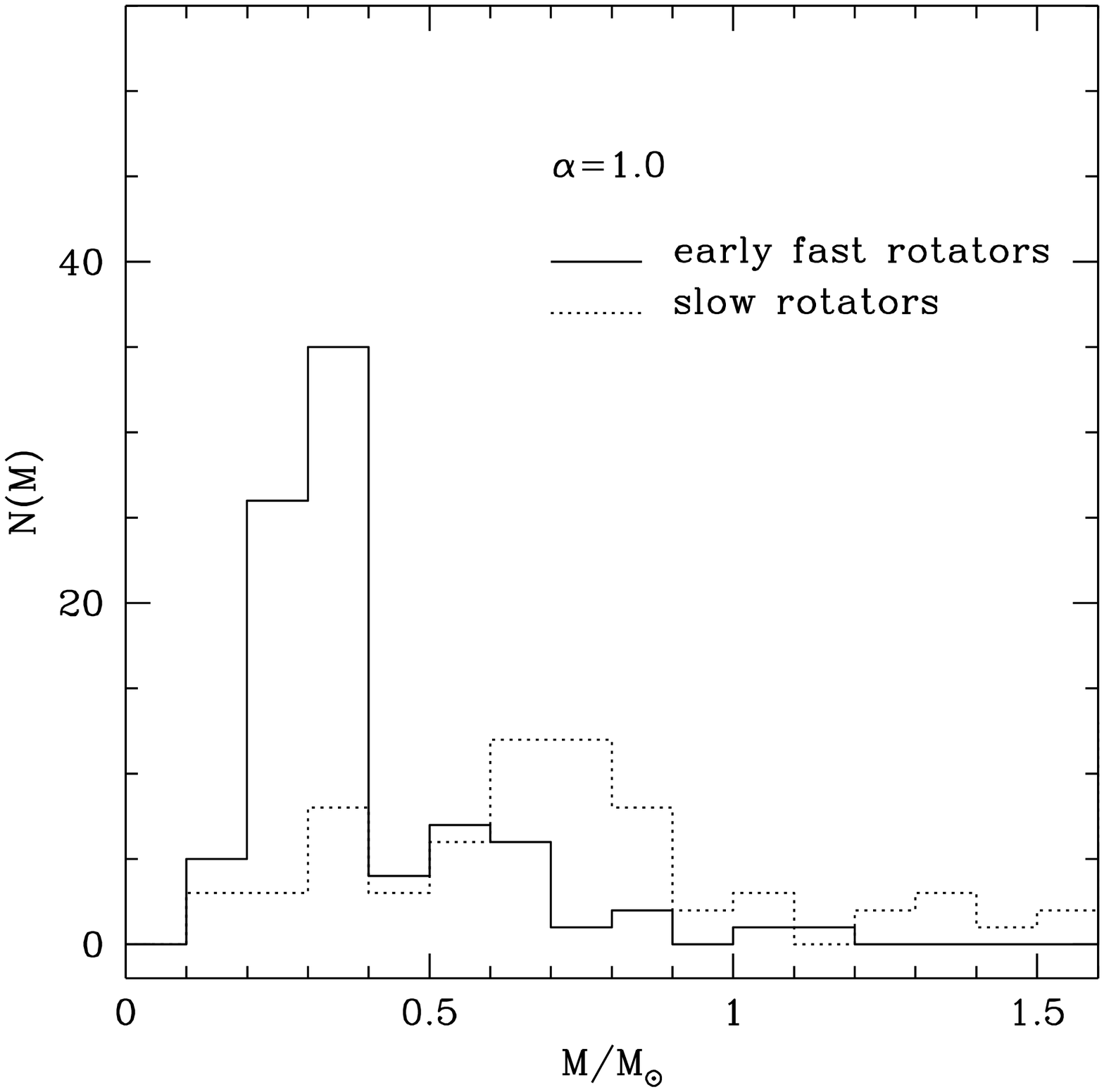}
\includegraphics[width=5.9cm]{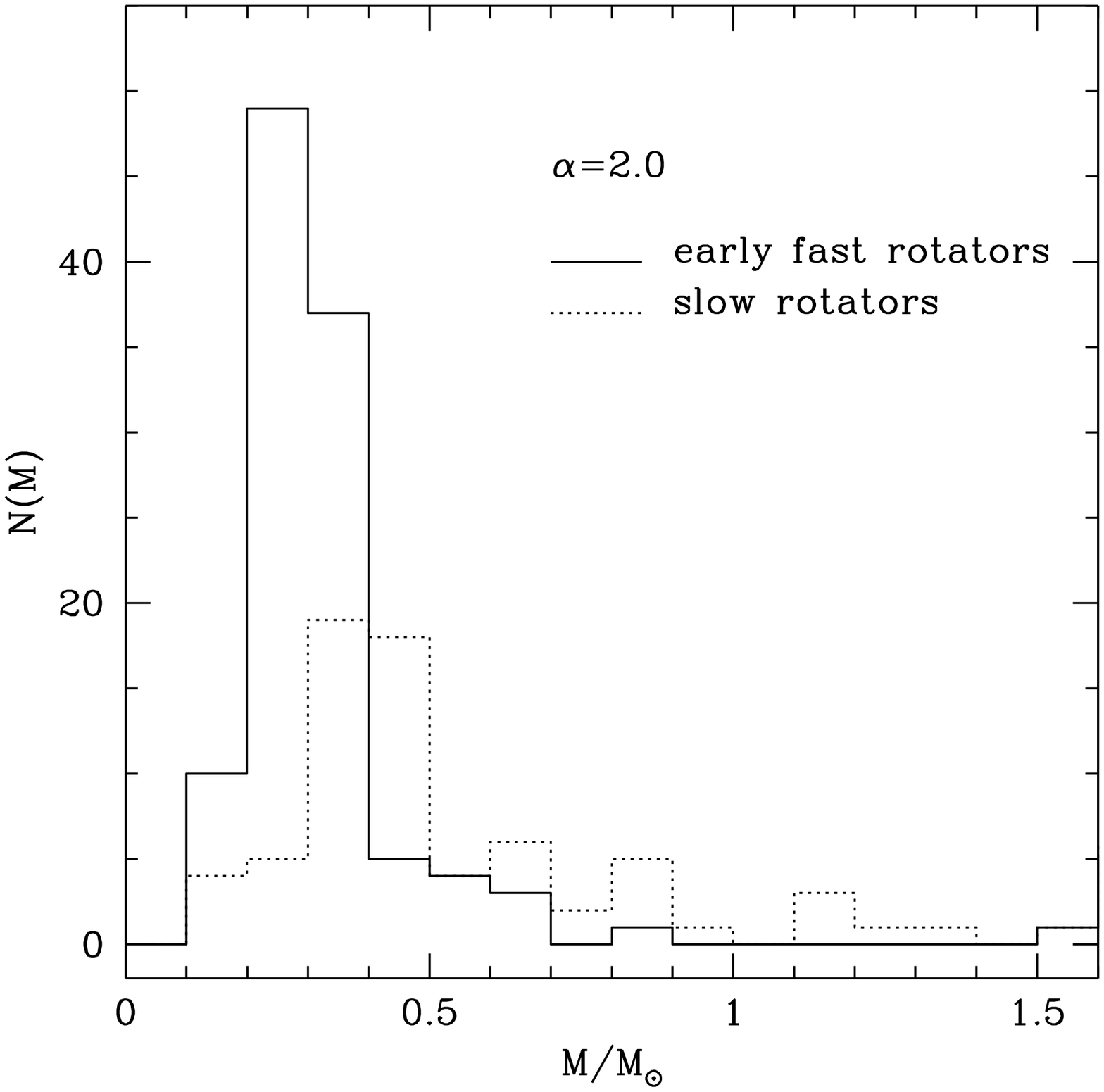}
\includegraphics[width=5.9cm]{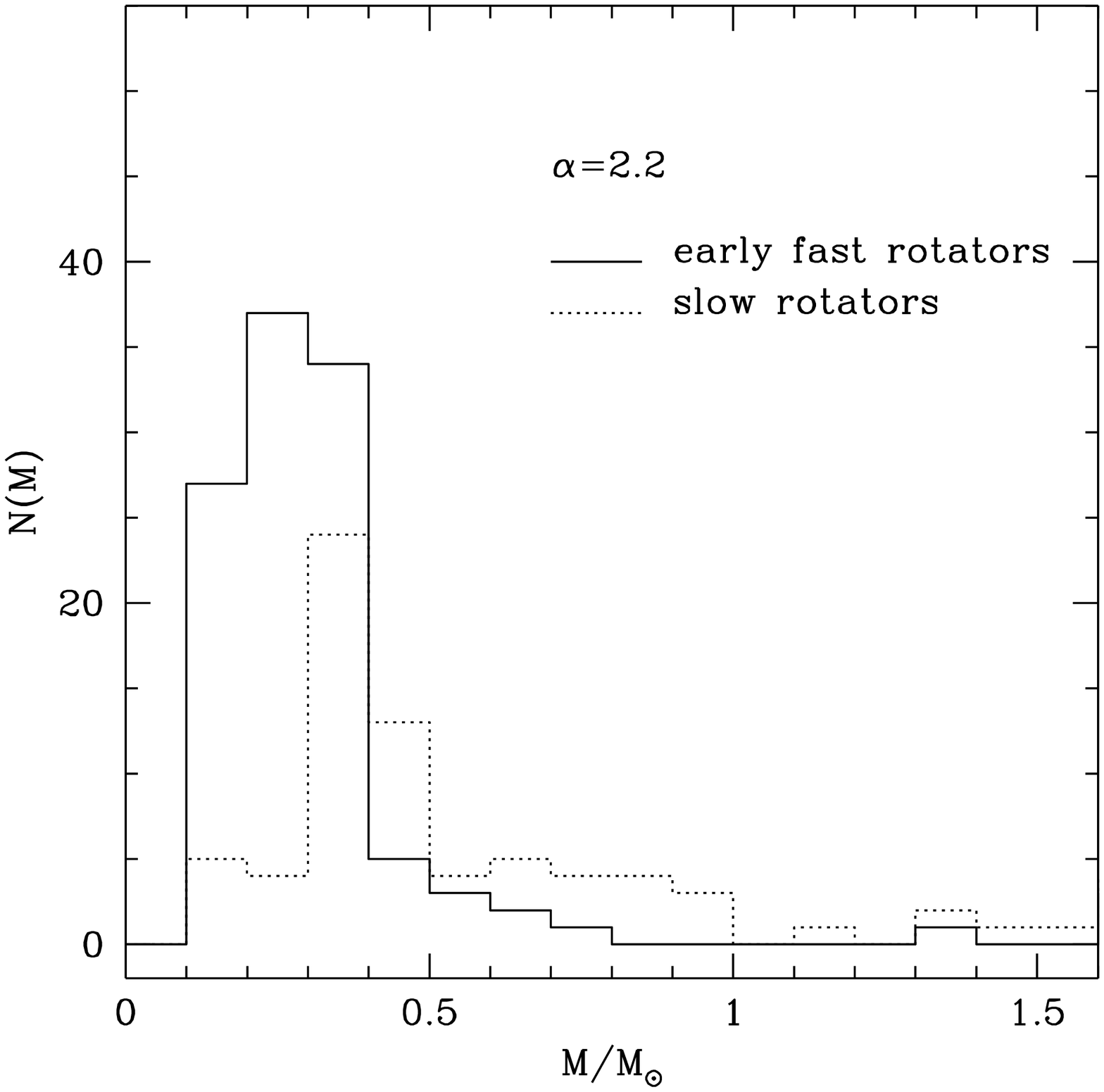}
}
\caption{
The mass distribution of the sources which according
to our analysis rotate fast since early evolutionary phases (solid lines) 
compared to those that are slow rotators in the current epoch (dotted lines).
This comparison is shown for the three sets of models ($\alpha1.0$, $\alpha2.0$
and $\alpha2.2$). 
}
         \label{histodisk}%
\end{figure*}

This dichotomy indicates that either (i) disk locking is responsible for the 
presence of the secondary peak, and stars with M$>$M$_{\rm tr}$ tend to be embedded 
in their disk longer than their low mass counterparts, 
(ii) the locking time is similar,
but the masses $>$M$_{tr}$ evolve faster and a larger fraction of their pre--MS 
lifetime is locked,
or (iii) the ``locking period'' of the group with M$<$M$_{\rm tr}$ 
is significantly lower than $\sim$8d.
Our analysis confirms possible interpretations of the observed distribution of 
periods given by Herbst et al.\ (\cite{herbst02}) (see their Fig.~15). The 
uncertainty on the convection model simply alters the transition mass from 
$\sim$0.25M$_{\odot}$ to a maximum of 0.5M$_{\odot}$ for LCE models. 

\begin{figure*}[htb]
\centering {
\includegraphics[width=5.9cm]{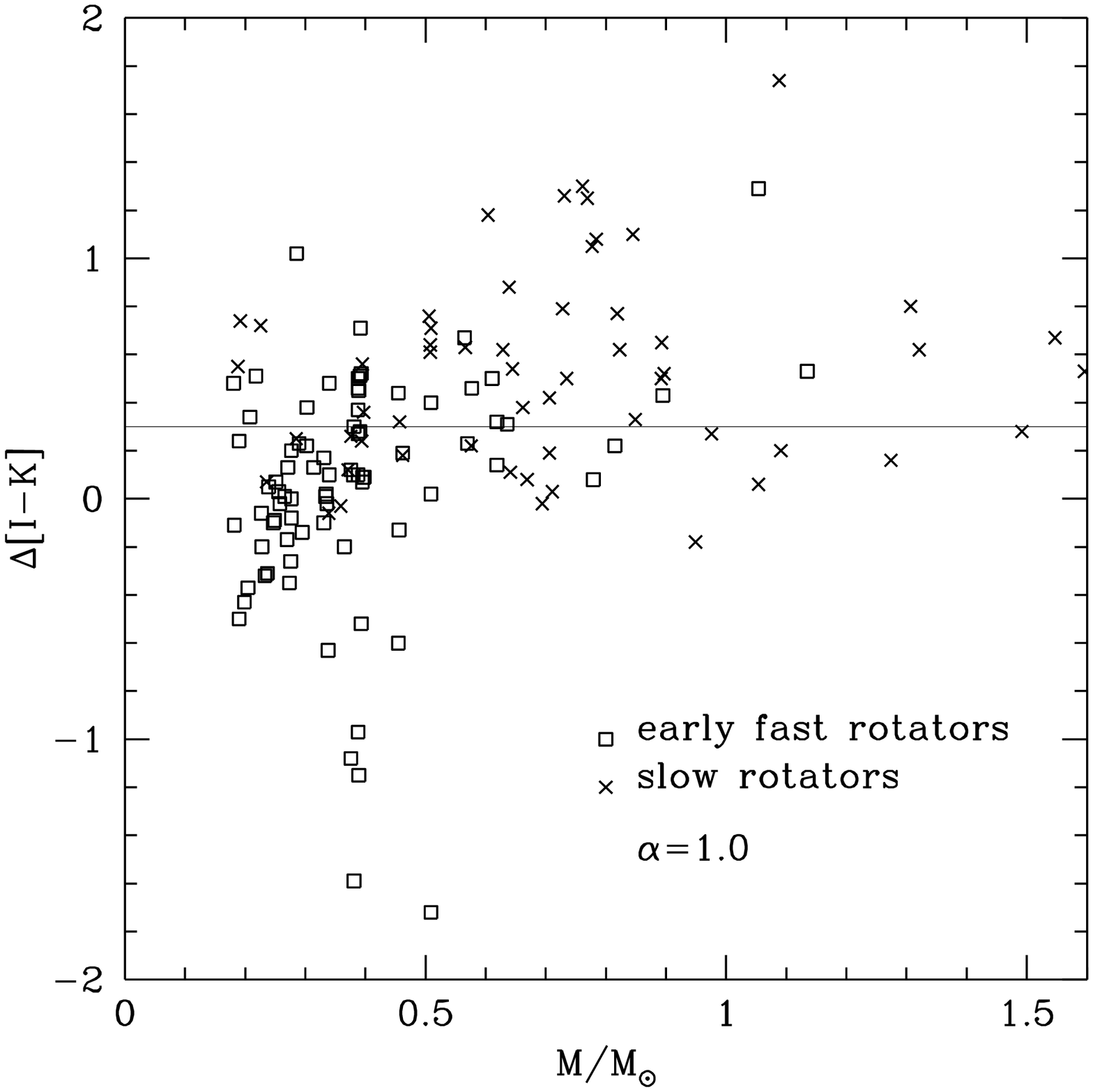}
\includegraphics[width=5.9cm]{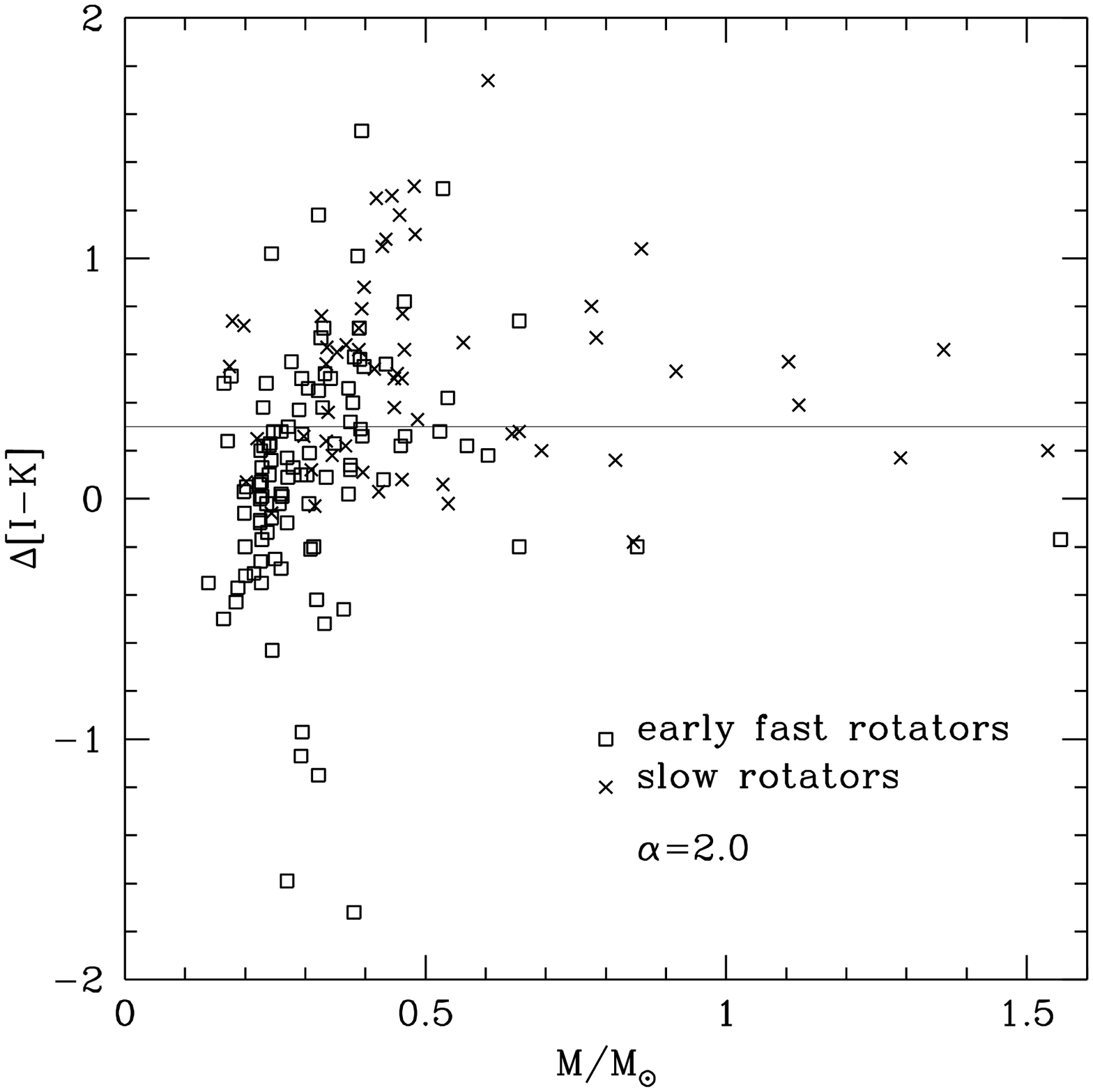}
\includegraphics[width=5.9cm]{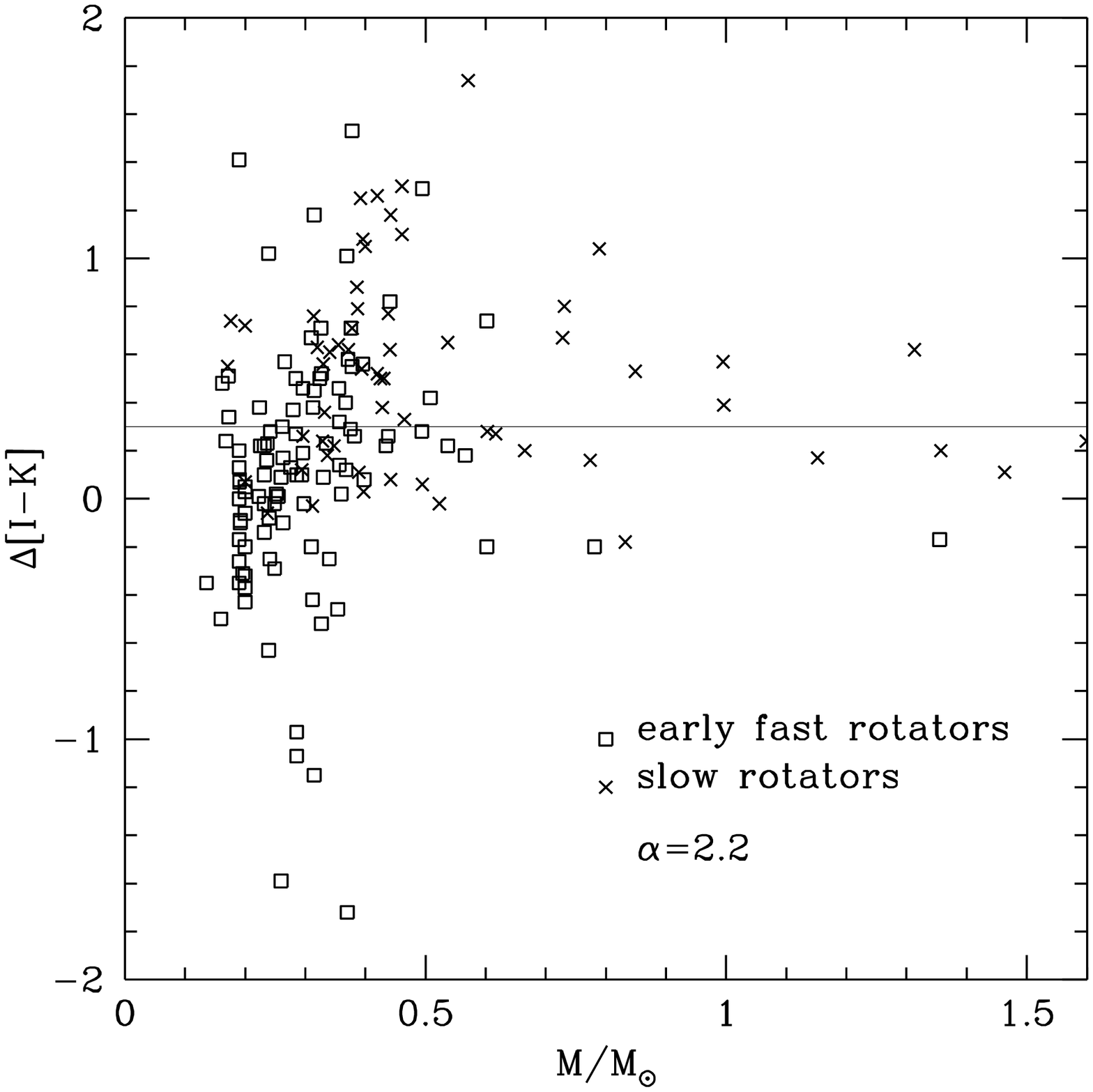}
}
\caption{
The observed infrared excess of our sample stars plotted against their inferred
mass according to the three sets of models used in this work.} 
         \label{excess}%
\end{figure*}

\subsection{Disk locking and the disk lifetime}
Following the suggestion by Herbst et al.\ 
(\cite{herbst02}), that the longer period peak in the distribution 
indicates that some stars are locked in their disks with a period near
8 days,
we simply considered the stars with periods larger than a threshold 
period (P$_{\rm thresh}$), which we put at 8d, as locked. For stars
at P$<$P$_{\rm thresh}$, unlocked stars according to our criterion,
we determine the 
epoch at which their period was equal to 8 days. This would be the 
time at which the stars would have lost their 
disks, and began a constant angular momentum evolution.
Here we consider that angular momentum losses by magnetic braking
are negligible at the pre--MS, since its
timescale is much longer than the evolutionary timescale during pre--MS
evolution.
The temporal variations of radius and angular velocity were determined on 
the basis of our tracks
once the mass was assigned. Following this hypothesis, we found some 
stars which had P=P$_{\rm thresh}$ at an age younger than 
$10^5$yr. In our interpretation, these stars have lost their 
disks very early and can be considered to have evolved without a disk.
In this way, we identify three distinct populations:
{\vspace{-.5\baselineskip}
\begin{enumerate}
\item{early fast rotators -- stars locked only for ages 
$<10^5$yr,}
\item{slow rotators -- stars probably still disk embedded,}
\item{moderate rotators.}
\end{enumerate}
}

Stars in the last group may have lost their disks at ages greater than 
$10^5$yr. They represent, on average, $\sim$45\% of the stars of the 
whole sample and this percentage do not change significantly with the choice of
$\alpha$.
As long as the assumption that the locking period is independent of 
mass is valid, the disks seem to survive longer for higher masses. For all sets 
of tracks, the percentage of early fast rotators is $\sim$40\% for 
M$<$0.4M$_{\odot}$, and gradually drops below $\sim$12\% for 
M$>$0.6M$_{\odot}$.  

The mass distribution of the slow and early fast rotators is shown
in Fig.~\ref{histodisk}. We note, in particular, the maximum at 
0.6$-$0.8$\,M_{\odot}$ (for LCE models) and 0.3$-$0.5$\,M_{\odot}$ 
(for HCE models) characterizing the slow rotator population 
(dotted line).
We can compare the fraction of early fast rotators identified in the ONC with 
the non-accreting fraction of TT and Brown Dwarfs in star forming regions 
of similar ages ($\rho$ Oph and Taurus, Mohanty et al. \cite{mohanty05}). The lower limit 
to this non-accreting fraction is $\sim$35\%, not very different from our 
result.

In our analysis we use the rotation period as an indicator of the
presence of a disk surrounding these stars. 
In order to test the reliability of this hypothesis, we
should use several observational indicators of the presence of disk
and accretion, like the infrared excess $\Delta$[I$-$K], the 
equivalent width of \ion{Ca}{II} line, the excess in the L-band and 
H$_{\alpha}$, Ca, O emission lines.
Near IR excess as disk indicators and EW \ion{Ca}{II} as accretion
diagnostic must be used with caution (Hillenbrand \cite{hillen97}). 
But, as this is mainly a 
theoretical work, we will check what is already in the literature for 
ONC stars, namely  IR excess $\Delta$[I$-$K] and EW \ion{Ca}{II}, 
mainly as additional arguments.   

It is expected that
still locked stars have $\Delta$[I$-$K]$>$0.3, and those that evolved 
without disk should have infrared excess significantly
lower than this threshold value (Herbst et al.\ \cite{herbst02}). 
We report in Fig.~\ref{excess} the observed stars on the plane  
$\Delta$[I$-$K] vs.\ mass. We can see that sources that we identified as 
still locked (slow rotators - crosses) are mainly concentrated above the $\Delta$[I$-$K]=0.3 
line, while those that evolved without a disk (early fast rotators - open squares) 
lie
mainly below it, for the three sets of models. This straight correlation between the infrared 
excess and our derivation of the presence of a disk agrees with our
theoretical considerations.

The equivalent width of \ion{Ca}{II} lines is commonly used as an indicator of
an active accretion process. For accreting stars we expect emission lines
and EW(\ion{Ca}{II})$<$$-$1, while for non-accreting objects we have EW(\ion{Ca}{II})$>$1
(Flaccomio et al.\ \cite{flacco03b}).
We cannot expect a direct correlation between $\Delta$[I$-$K] and 
EW(\ion{Ca}{II}), because some stars might still have a disk, although no
longer accreting. Yet, we expect that the observed stars with
EW(\ion{Ca}{II})$<$$-$1 and $\Delta$[I$-$K]$>$0.3 should have a disk surrounding them,
and the disk locking mechanism should be active. We could identify
$\sim$40 stars with known rotational periods that satisfy both requests, 
about 30\% of which are actually identified as still disk embedded according 
to our criterion ($P$$>$$P_{\rm thresh}$=8d).
Part of the remaining have long rotational periods, very close to 
$P_{\rm thresh}$, which suggest the presence of a disk.

\subsection{An alternative view: the role of the magnetic field}
The idea that disk-locking is responsible for the dichotomy observed, 
which eventually leads to the presence of two peaks in the 
distribution of periods of the ONC stars, was recently criticized by
Barnes (\cite{barnes}), who argued that the possible role of disks
can be only to set an initial distribution of periods, since disk-locking
should affect all the stars equally.

Based on the observed color-period diagrams of several
open clusters (e.g. IC 2391, $\alpha$ Per, Pleiades, M34, etc.), Barnes 
(\cite{barnes}) found
that a double population with distinct rotational properties   
characterizes any stellar association. Further, he found that a systematic trend with 
age is apparent, namely that older clusters have a smaller number of rapid rotators, that
eventually disappear at ages $\sim$800 Myr.

\begin{figure*}[htb]
\centering{
\includegraphics[width=5.9cm]{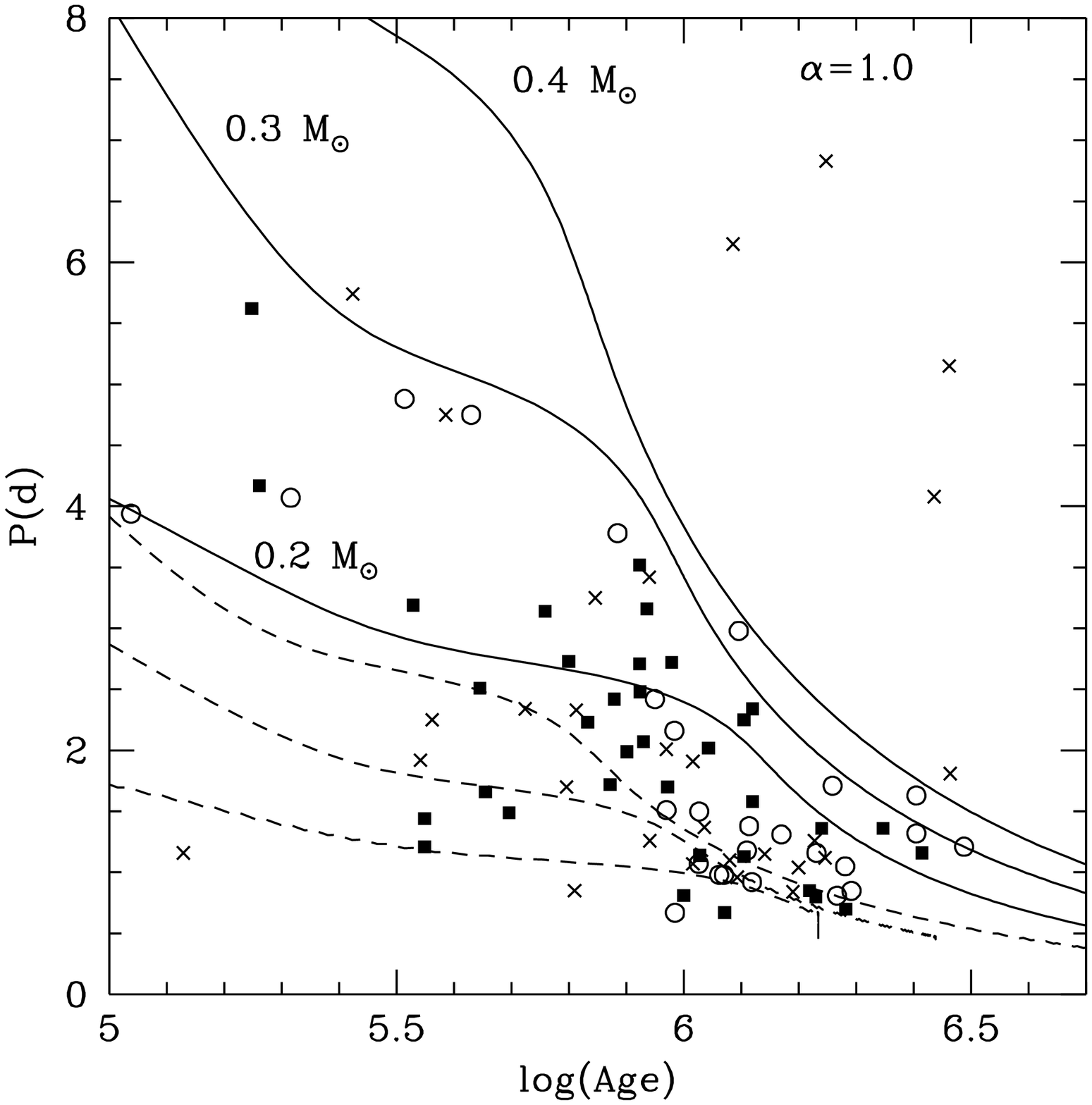}
\includegraphics[width=5.9cm]{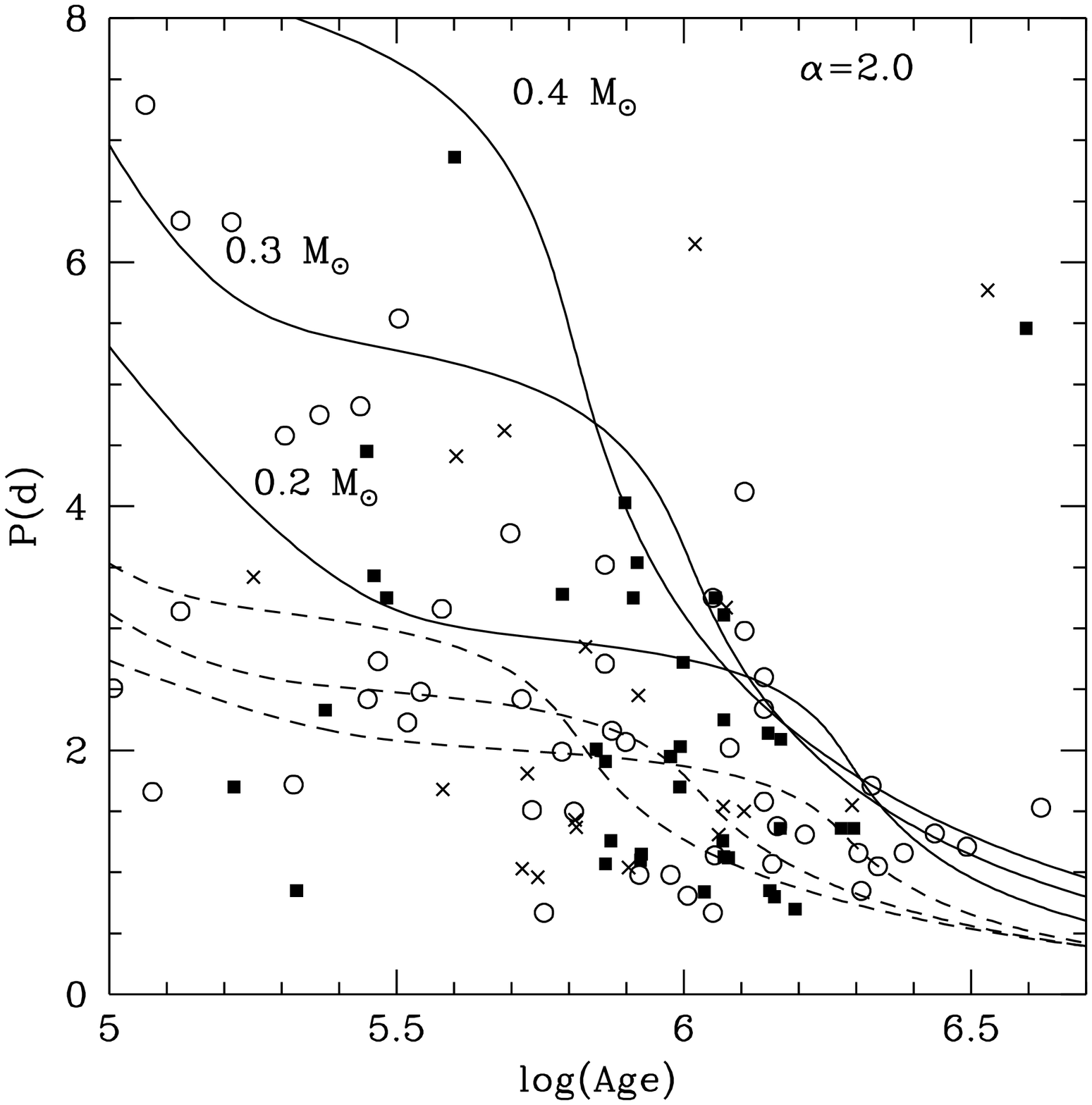}
\includegraphics[width=5.9cm]{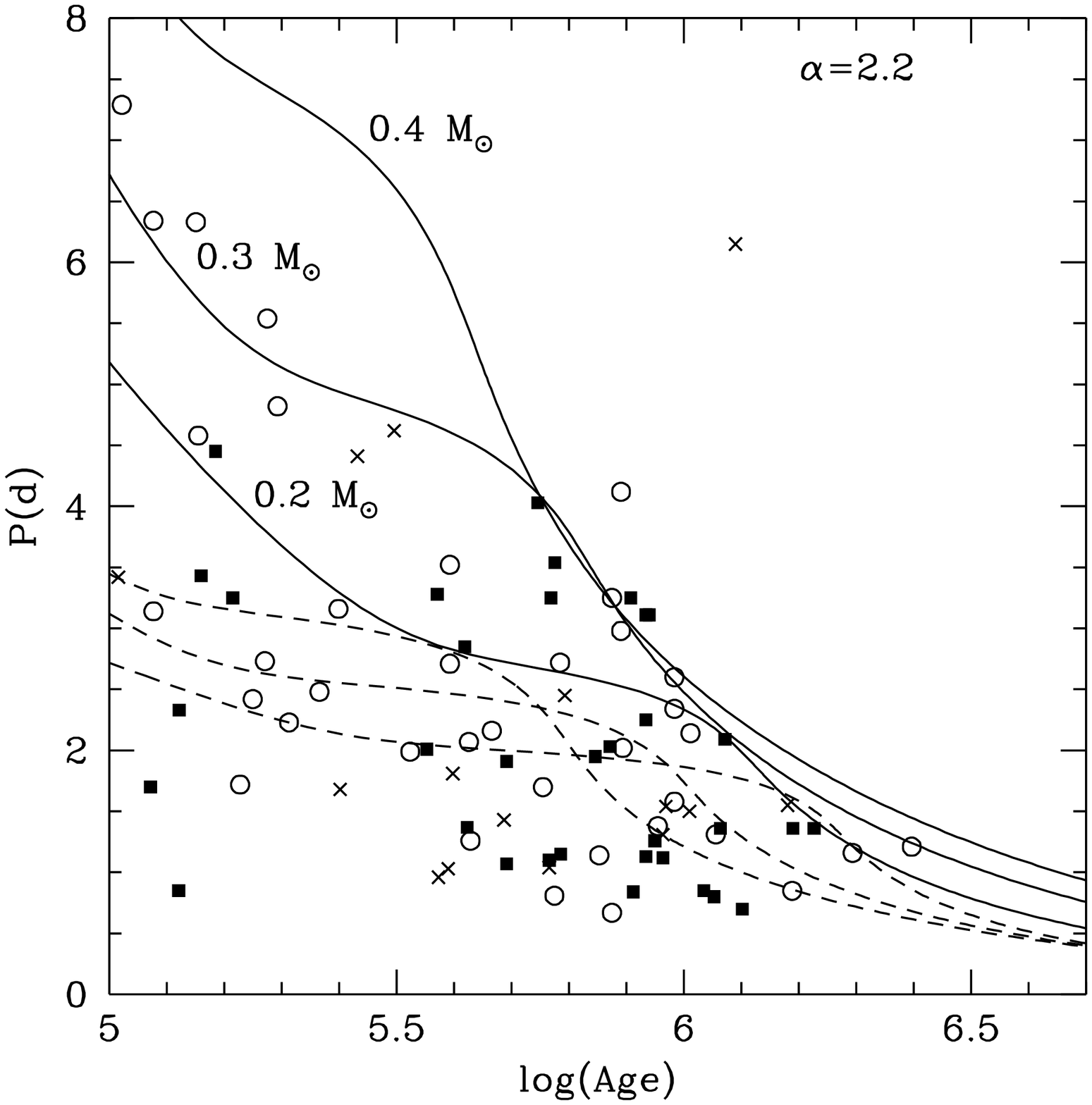}
           }
\caption{The temporal evolution of the periods of our
three sets of models with mass in the range
0.2$\,M_{\odot}$$\leq$$M$$\leq$0.4$\,M_{\odot}$ evolved
starting with an initial
angular momentum calculated on the basis of the Kawaler
(\cite{kawaler}) prescription (solid lines), and with the same values
multiplied by a factor of 3 (dashed lines). The stars that we 
suppose to be evolved with a constant angular momentum, i.e., without a 
disk, are also shown. The $\circ$ symbols identify the mass range
0.2$<$$M/M_{\odot}$$<$0.3, {\bull} is used for
0.3$<$$M/M_{\odot}$$<$0.4,
and $\times$ for all the other masses.
\label{evol}}
\end{figure*}

Barnes (\cite{barnes}) interpreted these observations as a result of a 
different morphology of the magnetic field configuration. According to his 
suggestion, rapid rotators have small scale magnetic fields associated with their 
convective region, which cannot be anchored either to the inner radiative core 
or to the star's external layers. All the fully convective stars should belong 
to this group. Conversely, the slowly rotating stars are characterized by large 
scale magnetic fields, probably associated with the presence of an interface 
dynamo between the external convective region and the internal radiative zone. 
In this case the process of spinning down the star is much more efficient. The 
dynamo, that is probably created by the decoupling between the convective and 
radiative zones, anchors the spun-down convective envelope to the rapidly 
rotating core, thus favouring a constant migration of the stars belonging to 
the rapidly rotating group to the slowly rotating sample. This should explain 
the complete absence of the fast rotator sequence at old ages (see Barnes 
\cite{barnes}, Fig.~1). Due to the young age of the ONC, both populations 
should be present there, as we find. 
This interpretation is not correct if we take our results at face value: if we rely on 
our attribution of masses and ages, the great majority of the stars observed 
are indeed fully convective, i.e. should all belong to the rapidly rotating 
sequence. We suggest that the magnetic field itself plays 
a role in inhibiting convection (Gough \& Tayler \cite{gough}; Moss 
\cite{moss}; Ventura et al.\ \cite{ventura}) and favours an earlier appearance 
of a radiative core in the stars having M$>$M$_{\rm tr}$. Consequently, we do not 
reject the idea that the presence of a double population is indeed due to the 
magnetic field configuration, rather than to the effects of a disk-locking 
mechanism.  

\subsection{A constant angular momentum evolution?}
\label{sec-evol}

\begin{figure*}[htb]
\centering{
\includegraphics[width=5.9cm]{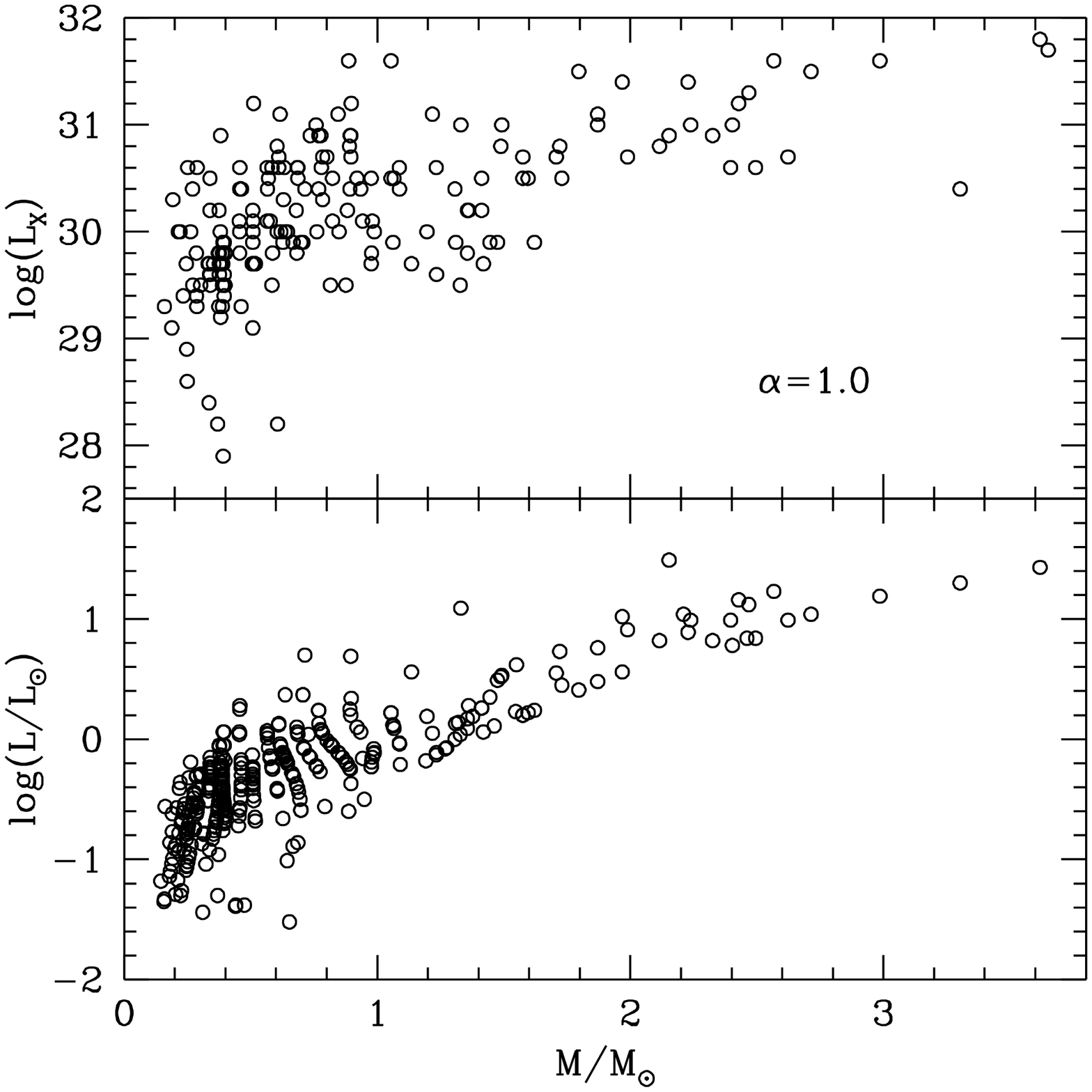}
\includegraphics[width=5.9cm]{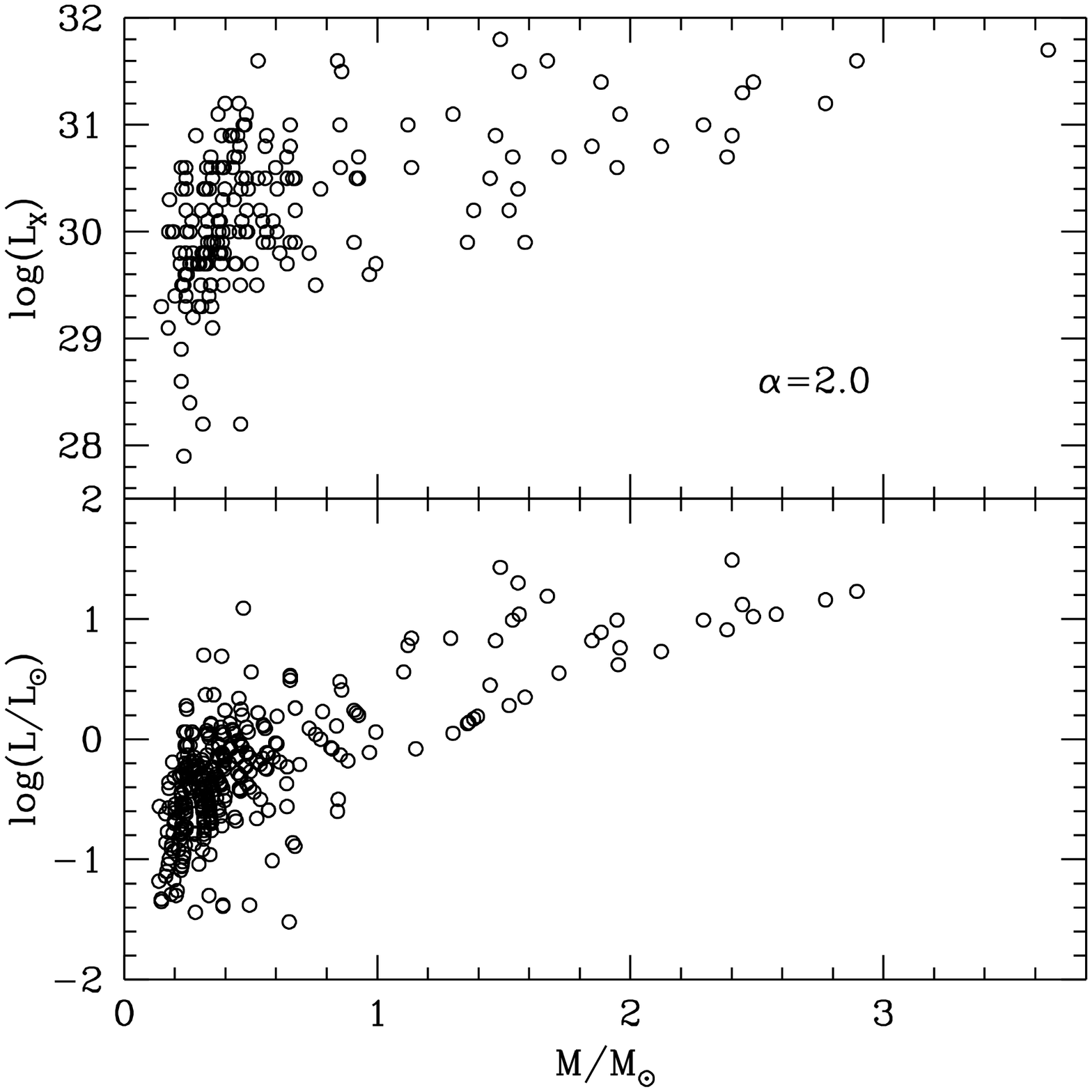}
\includegraphics[width=5.9cm]{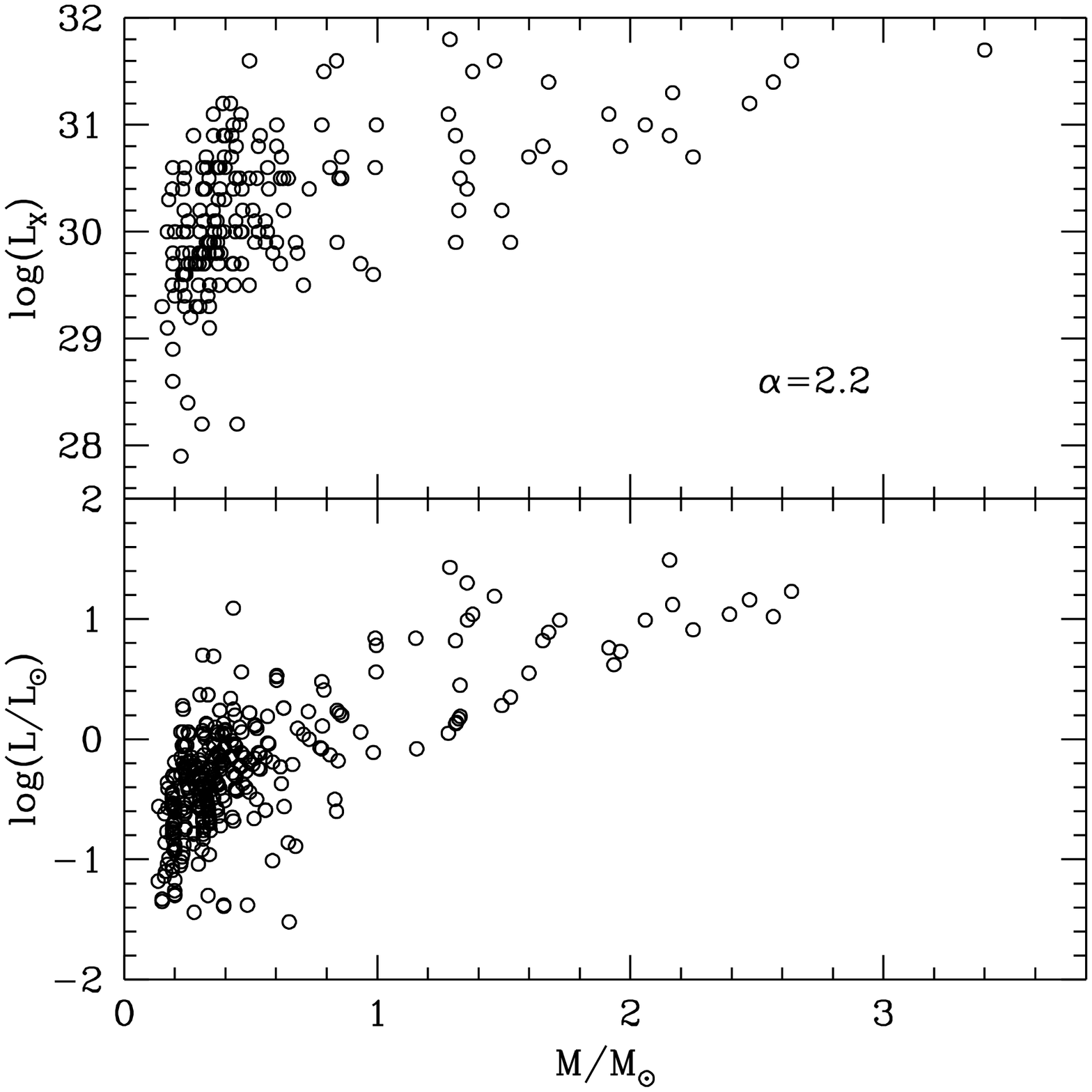}
          }
\caption{{\bf Top}: $L_{\rm X}$ luminosity
(ergs/s)
plotted against mass (obtained with our three sets of models) for the ONC 
observed sources with known rotational periods. {\bf Bottom}: 
The same as top, with bolometric luminosity on the vertical axis.
\label{lxlbol}}
\end{figure*}

For those stars that we identified to have lost the disk in their early 
evolutionary phases, i.e., the early fast rotators, we may test the hypothesis 
that they evolved at constant angular momentum from the beginning. The 
distribution of these objects in the $P$ vs.\ inferred age plane is shown in 
Fig.~\ref{evol} for $\alpha$=1.0, $\alpha$=2.0 and $\alpha$=2.2 models. We used 
different symbols as mass identifiers. We can see a clear trend towards shorter 
periods for older ages, especially for the $\alpha1.0$ models, possibly 
indicating angular momentum conservation. We checked the possibility of 
reproducing the observed rotational pattern with age by means of our rotating 
models. This approach allows us to find out the range of initial 
angular momenta that, for each mass, must be used to calculate the models. We 
limit this analysis to the subsample of the mass range given in Table 
\ref{results}. We 
divided the observed sources into three classes of mass, indicated in 
Fig.~\ref{evol} with open circles masses in the interval 0.2$<$$M/M_{\odot}$$<$0.3, with full 
squares 0.3$<$$M/M_{\odot}$$<$0.4 and with crosses all the remaining. 
The solid lines indicate the temporal variation of the rotational periods 
according to the evolution of our three sets of models with masses $M$=0.2, 
$0.3$, 0.4$\,M_{\odot}$ calculated by assuming an initial angular momentum 
following the prescriptions given in Eq.~\ref{kaweq}. We note that the temporal 
evolution of the periods vary with the parameter $\alpha$, as it affects the 
radius of the stars. These curves can only reproduce the upper envelope of the 
observed loci, but, particularly at the ages shared by the bulk of the ONC 
stellar population, they lead to rotational velocities too slow with respect to 
most of the observed values. 

To reproduce the rotation period of the fastest stars, we need to use an 
initial angular momentum at least three times larger than that prescribed by 
Kawaler (\cite{kawaler}), if we use LCE models, and even larger for HCE models. 
To fully bracket the observed periods it is necessary to assume a distribution 
of initial angular momenta $J_{\rm in}$, at least, in the range $J_{\rm 
kaw}$$<$$J_{\rm in}$$<$3$J_{\rm kaw}$. This result can be used to extend the 
Kawaler (\cite{kawaler}) prescription to the very low mass stars. 

\subsection{The X-ray emission of the ONC stars}
Flaccomio et al.\ (\cite{flacco03a}, \cite{flacco03b}) and Stassun et al.\ 
(\cite{stassun04}) performed deep analyses of the archival {\it Chandra} data 
and derived the $L_{\rm X}$ luminosity of all the sources included in the 
Hillenbrand (\cite{hillen97}) sample. Their goal was to correlate $L_{\rm X}$ 
of pre-MS stars with the factors most likely driving the X-ray emission 
itself, i.e., accretion and rotation. Their main finding was the lack of a 
clear correlation between $L_{\rm X}$ and rotational period. They interpreted 
this result as evidence that the ONC pre-MS stars are indeed in the 
``super-saturated'' regime of the rotation-activity relationship. This seems to 
be confirmed by the average value of the fractional X-ray luminosity 
$\log(L_{\rm X}/L_{\rm bol})$$\sim$$-$3.6, that is slightly smaller than the 
main sequence saturation value of $\log(L_{\rm X}/L_{\rm bol})$$\sim$$-$3.

Concerning the relationship between accretion and X-ray emission, Stassun et 
al.\ (\cite{stassun04}) found that accreting stars have X-ray luminosities on 
average lower than their non-accreting counterparts, as a possible result 
of X-ray extinction by circumstellar gas in magnetospheric accretion columns. 
We could not find any clear
correlation between EW(\ion{Ca}{II}) and $L_{\rm X}$, although this might be a
consequence of the smaller sample of stars with evidence of accretion
(i.e.\ EW(\ion{Ca}{II})$<$$-$1) used 
in the present work, compared to the complete sample by Hillenbrand 
(\cite{hillen97}) and Hillenbrand et al.\ (\cite{hillen98}) analyzed by 
Stassun et al.\ (\cite{stassun04}). 

\begin{figure*}[htb]
\centering{
\includegraphics[width=5.9cm]{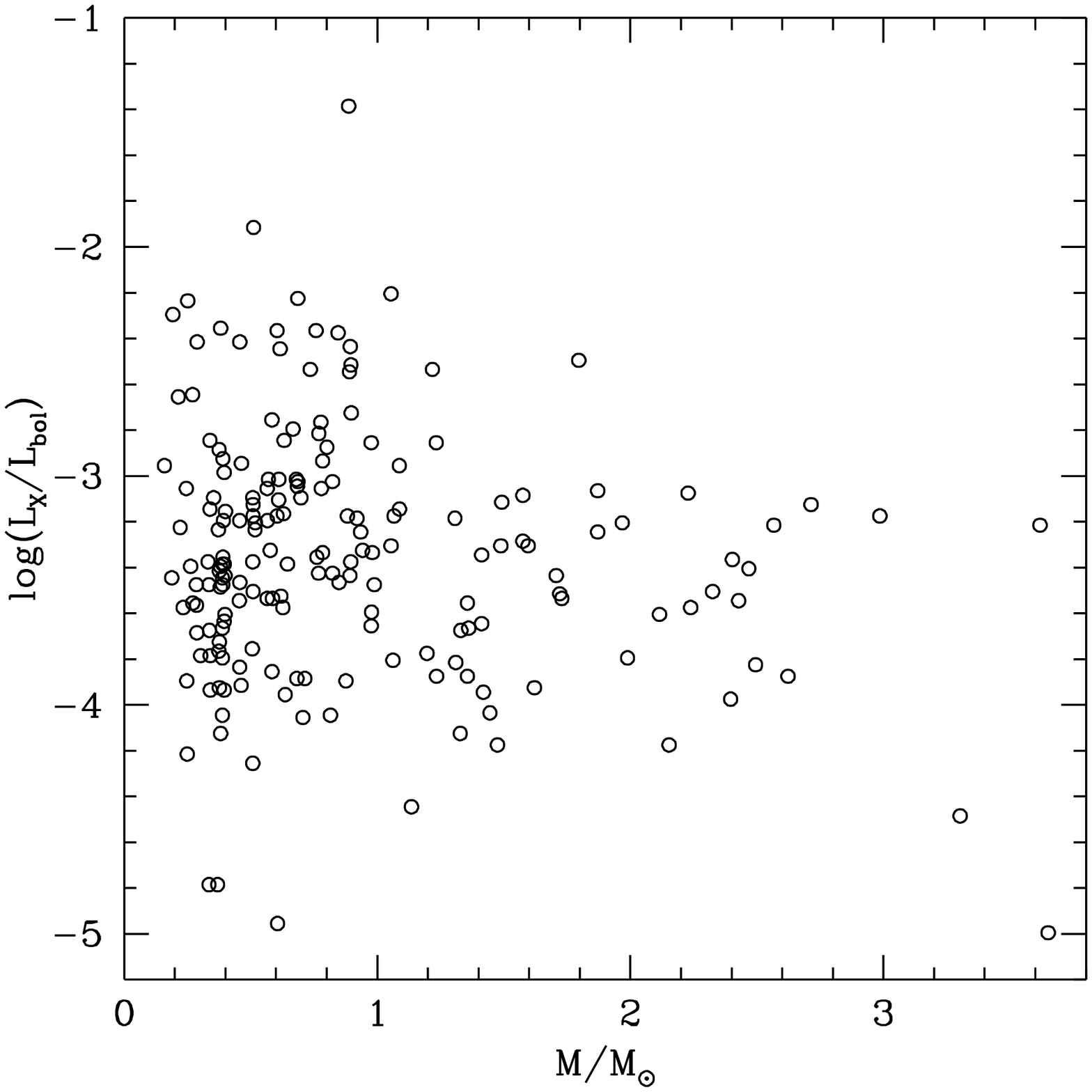}
\includegraphics[width=5.9cm]{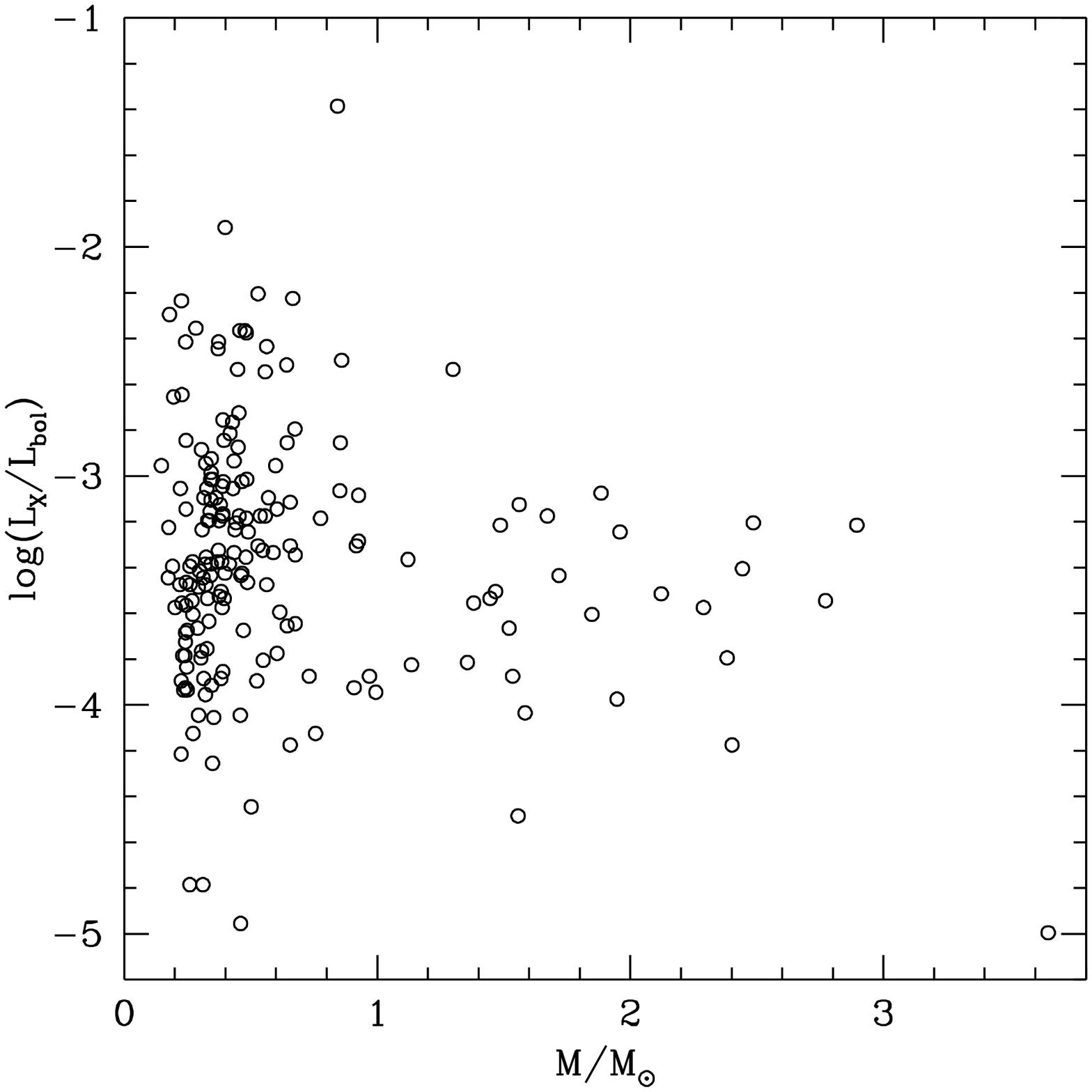}
\includegraphics[width=5.9cm]{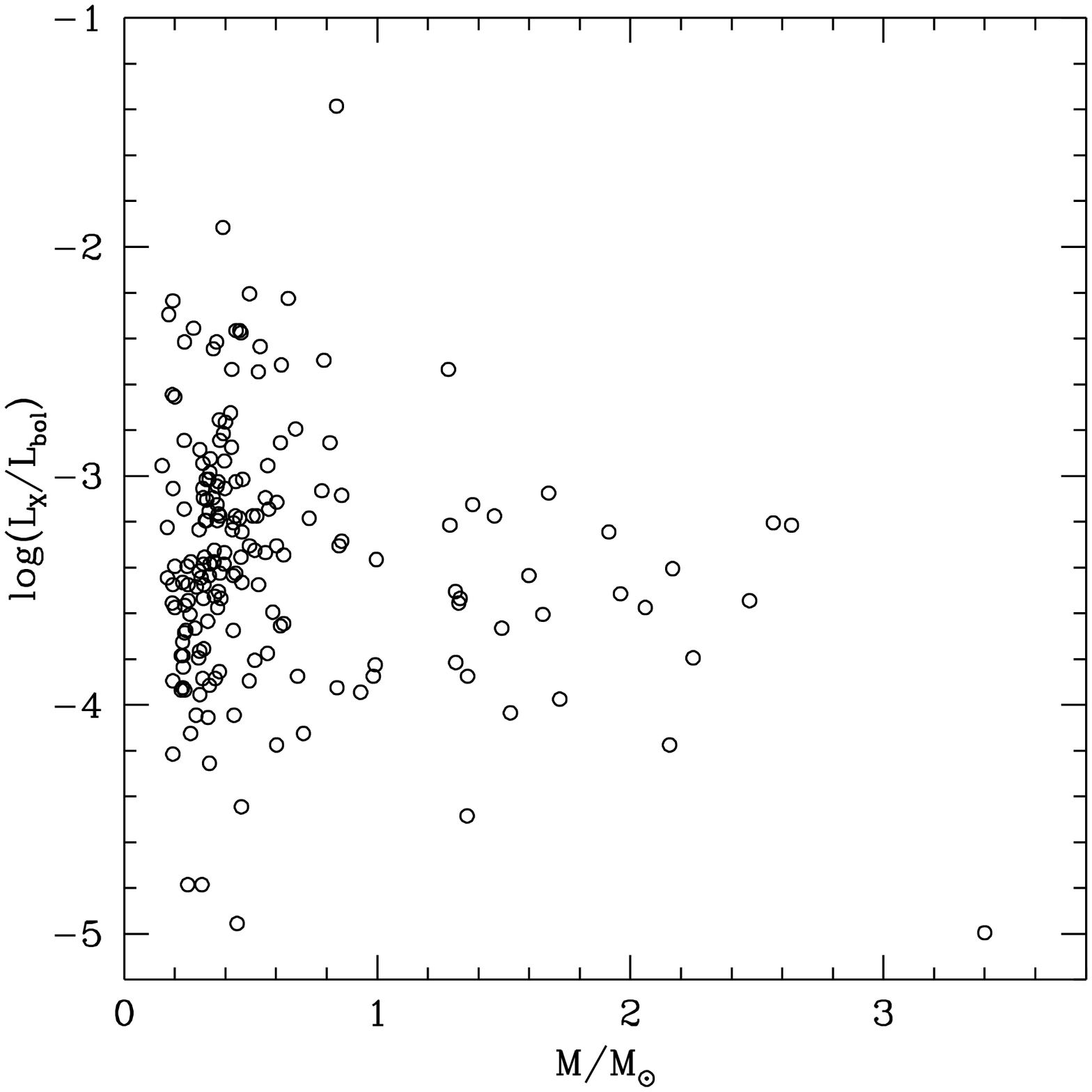}
          }
\caption{The fractional X-ray luminosity as a function of mass
         for the same stars and models shown in Fig.~\ref{lxlbol}.
         \label{lxlbrat}}
\end{figure*}

We used our determinations of mass to look for any relationship between stellar 
mass and X-ray luminosity. In agreement with Flaccomio et al.\ 
(\cite{flacco03a}), we find that $L_{\rm X}$ is correlated with mass (upper 
panels of Fig.~\ref{lxlbol}). This trend is not due to a qualitative difference 
in the X-ray emissions, but rather to a general correlation between mass and 
bolometric luminosity (lower panels of Fig.~\ref{lxlbol}). This is confirmed in 
Fig.~\ref{lxlbrat}, where we see that the $L_{\rm X}/L_{\rm bol}$ ratio is 
practically independent of mass for the three $\alpha$ models we have used. We 
note the high dispersion around the average value of $\log(L_{\rm X}/L_{\rm 
bol})$$\sim$$-$3.6, present at the lowest masses, that is probably connected to 
the lower luminosities of these objects.

\section{Conclusions}
\label{sect8}
We use the observed stellar population of the ONC to test our pre-MS 
evolutionary tracks, to better understand the 
main physical properties characterizing the evolution of young stars. By 
comparing the location of the tracks in the HR diagram with the position of the 
observed objects, we assign to any single star a mass and an age, for three 
different convection efficient models. 

We find the well known result that the treatment of convection is generally the 
most relevant physical input in determining the T$_{\rm eff}$ of pre-MS tracks 
in the HR diagram. The boundary conditions adopted also play a non-negligible 
role in determining the path followed by the evolutionary sequences on the HR 
plane. Gray models are systematically hotter than their non-gray counterparts. 
This effect, for the range of masses and ages at which most of the ONC 
population is found, has a similar quantitative effect as a change of the 
convective model. The use of non-gray models is recommended to describe these 
early evolutionary phases.

On the observational side, we find that the bulk of the observed stars in the 
ONC have masses in the range 0.2M$_{\odot}$$\leq$M$\leq$0.4M$_{\odot}$, for all 
non-gray models. The age distributions are more affected by the choice of the 
MLT parameter
$\alpha$. Ages are 1$-$2 Myr from the $\alpha1.0$ set, 0.6$-$2.5Myr for 
$\alpha2.0$ set and 0.4$-$1.6Myr for $\alpha$2.2 models. This study confirms 
the presence of a dichotomy in the rotational properties between the objects 
with M$<$M$_{\rm tr}$, whose period distribution peaks at short values, and stars 
with M$>$M$_{\rm tr}$, that present a secondary peak at $P$$\sim$8d. The transition 
value of mass between the two populations is at M$\sim$0.5M$_{\odot}$ for LCE 
and at M$\sim$0.35M$_{\odot}$ for HCE models. If disk-locking is  
responsible for the secondary peak observed in the overall period distribution, 
these results can be interpreted by assuming either that the masses M$<$M$_{\rm tr}$ 
lose their disk earlier, or that their locking period is shorter. The X-ray 
emission shows no correlation with period, supporting the suggestion
that these stars are 
indeed in the super-saturated regime of the rotation-X ray luminosity 
relationship. The correlation of the X-ray flux with mass appears to be the 
consequence of the increased average luminosity of more massive objects.

For the low-mass stars that presumably evolved without a disk, we find that
our results are consistent with an evolution-conserving angular momentum. 
The comparison between the model period evolution and the
observed values suggests that initial angular momenta at least 3 times larger than
those found by means of the Kawaler (\cite{kawaler}) law are needed for stars with
mass in the range 0.2$\,M_{\odot}$$\leq$$M$$\leq$0.4$\,M_{\odot}$. This analysis
was not possible for the higher masses sample, due to the lack of a
statistically meaningful sample of rapidly rotating stars.

The idea that the double population of the ONC can be explained on the basis of 
a different morphology of the stellar magnetic fields seems to be ruled out by 
%the fact the almost all the observed sources are fully convective, according to 
the fact that almost all the observed sources are fully convective, according to 
our interpretation. However, a mechanism (e.g. the magnetic field itself) 
inhibiting convection might favour an earlier appearance of the radiative core, 
at least in some of the stars. In our analysis we found other 
indications that convection in the pre-MS may be affected by other parameter(s):
although 2D hydrodynamic simulations predict HCE in the pre-MS, we find two results 
in favour of LCE: (1) the lithium depletion in HCE models is too large to be 
consistent with the pre-MS depletion shown in young open clusters; (2) the
age distribution derived from HCE models for two groups of smaller and higher 
masses is very different. It may be that the lower convection efficiency
needed in the pre-MS is due to the structural effect of the dynamo-induced
magnetic field, as suggested by Ventura et al. (\cite{ventura}) and D'Antona et al.
(\cite{dantonaetal2000}). This possibility leads us not to dismiss the idea of an earlier appearance
of a radiative core in the M$>$M$_{\rm tr}$ group.

The evolutionary tracks (from 0.085 to 1.6M$_{\odot}$) and isochrones 
(from 2$\times$10$^5$ to 1$\times$10$^7$Myr) are available from the following web site:

www.mporzio.astro.it/$\sim$tsa

\begin{acknowledgements}
The authors thank Dr. Keivan Stassun for providing the
observational data of ONC. Nat\'alia R. Landin would like to thank 
the Osservatorio Astronomico di Roma for their gracious hospitality 
during an extended visit. NRL and LPRV gratefully acknowledge financial
support from the Brazilian agencies CAPES, CNPq and FAPEMIG.
\end{acknowledgements}

\end{document}